\documentclass[aps,pra,twocolumn,showpacs,superscriptaddress]{revtex4}
\usepackage{graphicx}
\usepackage{amsmath}
\usepackage{amssymb}
\usepackage{dcolumn}

\input{epsf}
\begin{document}

\title{Occupation numbers
of the harmonically trapped few-boson system}

\author{K. M. Daily}
\affiliation{Department of Physics and Astronomy,
Washington State University,
  Pullman, Washington 99164-2814, USA}
\author{X. Y. Yin}
\affiliation{Department of Physics and Astronomy,
Washington State University,
  Pullman, Washington 99164-2814, USA}
\author{D. Blume}
\affiliation{Department of Physics and Astronomy,
Washington State University,
  Pullman, Washington 99164-2814, USA}

\date{\today}

\begin{abstract} 
We consider a harmonically trapped 
dilute $N$-boson system described by a low-energy
Hamiltonian with
pairwise interactions.
We determine the 
condensate fraction, defined in terms of 
the largest occupation number, of the
weakly-interacting $N$-boson system ($N \ge 2$)
by employing a perturbative treatment
within the framework of second quantization.
The one-body density matrix 
and the corresponding occupation numbers are compared 
with those obtained by solving the two-body problem
with zero-range interactions 
exactly.
Our expressions are also compared with high precision 
{\em{ab initio}} calculations for 
Bose gases with 
$N=2-4$ that interact through finite-range two-body 
model potentials. Non-universal corrections are identified
to enter at subleading order,
confirming that different low-energy Hamiltonians,
constructed to yield the same
energy, may yield different occupation numbers.
Lastly, we consider the strongly-interacting
three-boson system under spherically
symmetric harmonic
confinement and determine its occupation numbers 
as a function of the three-body ``Efimov parameter''.
\end{abstract}


\maketitle

\section{Introduction}
The weakly-interacting homogeneous Bose gas
has been studied extensively in the 
literature~\cite{bogo47,lee57,lee57a,huan57,wu59,huge57}.
Most commonly, the equation of
state of the homogeneous Bose gas
is expressed in terms of the square root of the 
dimensionless gas parameter 
$\rho [a_s(0)]^3$, where $\rho$ denotes the density and $a_s(0)$
the zero-energy 
$s$-wave scattering length. The leading order term
is the mean-field energy and the lowest order correction accounts
for
quantum fluctuations.
In an alternative approach~\cite{huan57,bean07,detm08,tan08}, 
the ground state energy 
of $N$ bosons  in a cubic box with periodic boundary
conditions 
has been
obtained by applying perturbation theory 
or the techniques of effective field
theory to the weakly-interacting regime.
As outlined by Lee, Huang and Yang~\cite{lee57a}, 
the latter approach
must reproduce
the equation of state of the weakly-interacting
homogeneous Bose gas if the energies of the ``subclusters''
are summed up carefully.

In addition to the energy, other observables 
of the homogeneous
weakly-interacting Bose gas have been considered.
The condensate fraction $N_0/N$, i.e., the fraction of particles
in the macroscopically occupied lowest momentum state, is a particularly
interesting quantity since it can be measured experimentally. Furthermore,
the connection between the condensate fraction and the superfluid
fraction has been investigated in 
the
literature, starting with the 
seminal works of London, 
Penrose and Onsager, and others~\cite{lond38,penr56,atkins,tilley}.
The condensate fraction has, as the energy per particle,
been expanded in
terms of the gas parameter $\rho [a_s(0)]^3$,
$N_0/N=1- 8/(3 \sqrt{\pi}) \sqrt{\rho [a_s(0)]^3} + \cdots$.
Application of the local density approximation shows that the
condensate fraction of the weakly-interacting Bose gas under spherically
symmetric harmonic confinement scales as 
$N/N_0=1 - 5\sqrt{\pi}/8 \sqrt{\rho(0) [a_s(0)]^3} + \cdots$, where
$\rho(0)$ denotes the peak density~\cite{dalf98}.

This work considers  $N$ identical
mass $m_a$ bosons under spherically
symmetric harmonic confinement with angular trapping frequency $\omega$.
For a review article
of trapped gases, the reader is referred to Ref.~\cite{blumereview}.
In the weakly-interacting regime, i.e., in the regime where the 
two-body $s$-wave scattering length $a_s(0)$
(expressed in units of $a_{\rm{ho}}$)
and the product of the two-body effective
range $r_e$ and $[a_s(0)]^2$ 
(expressed in units of $a_{\rm{ho}}^{3}$) are small, we determine 
expressions for the condensate fraction $N_0/N$;
here, $a_{\rm{ho}}$ denotes the harmonic oscillator length,
$a_{\rm{ho}}=\sqrt{\hbar/(m_a \omega)}$.
For trapped systems, the condensate fraction is related to the 
largest eigen value of the one-body density matrix.  
In particular, the largest eigen value or occupation number 
of the one-body density matrix defines the condensate fraction.
Our results are obtained by applying time-independent
perturbation theory
to the $N$-boson Hamiltonian
with pairwise 
zero-range interactions characterized by
$a_s(0)$ and $r_e[a_s(0)]^2$.
The perturbative expressions are compared with
highly 
accurate numerical results for Bose gases with $N=2-4$ that
interact through a sum of short-range two-body model potentials.
This comparison confirms that the leading-order term of the
condensate depletion scales as $(N-1)[a_s(0)]^2$.
At sub-leading order,
a 
non-universal correction appears, i.e.,
a correction which is independent of $a_s(0)$ and $r_e$
and which is not needed to reproduce the energy
of the finite-range system within an effective field theory 
approach~\cite{john09,john12}.

For the two- and three-boson systems, we go beyond the 
weakly-interacting regime.
For two 
harmonically trapped
bosons that interact through a regularized zero-range interaction
potential, we 
determine the occupation numbers as a 
function of the scattering length. For the three-boson system,
we consider the unitary regime 
[$1/a_s(0)=r_e=0$] and determine the occupation numbers as a 
function of the three-body phase or Efimov parameter.
The occupation numbers for the two- and three-body systems
show ``oscillatory behavior'' in the positive energy regime if plotted
as a function of the relative two-body energy and relative three-body
energy,
respectively. 
In the two-particle case, the
oscillations are associated with the fact that
the two-body $s$-wave phase shift 
changes by $2 \pi$ as the two-body energy changes by about
$2 \hbar \omega$.
In the three-particle case, in contrast, 
the
oscillations are associated with the fact that the three-body 
Efimov phase 
goes through cycles of
$2 \pi$ as the three-body
energy changes.

The remainder of this paper is organized as follows.
Section~\ref{sec_system} introduces the system Hamiltonian 
and defines the one-body density matrix and the occupation numbers.
Section~\ref{sec_twobody} discusses the occupation numbers of the
trapped
two-boson system.
Section~\ref{sec_nboson}
considers the weakly-interacting regime of the $N$-boson system.
Section~\ref{sec_threebody}
considers the strongly-interacting three-boson system.
Lastly, Sec.~\ref{sec_summary} concludes.
Mathematical details are relegated to Appendix~\ref{appendixA}
and Appendix~\ref{appendixB}.

\section{System Hamiltonian and definitions}
\label{sec_system}
We consider $N$ identical mass $m_a$
bosons that
interact through a short-range
interaction potential $V_{\rm{tb}}$
under external 
spherically symmetric harmonic confinement with angular trapping 
frequency $\omega$.
For this system, the Hamiltonian $H$ reads
\begin{eqnarray}
\label{eq_twobodyham}
H= 
\sum_{j=1}^N H_{\rm{ho}}({\bf r}_j)+ \sum_{j<k}^N V_{\rm{tb}}({\bf r}_{jk}),
\end{eqnarray}
where $H_{\rm{ho}}({\bf r}_j)$ denotes the single-particle harmonic
oscillator
Hamiltonian,
\begin{eqnarray}
H_{\rm{ho}}({\bf r}_j)=\frac{-\hbar^2}{2m_a} \nabla_{{\bf r}_j}^2 + 
\frac{1}{2} m_a \omega^2 {\bf r}_j^2,
\end{eqnarray}
and ${\bf r}_j$ the
position vector of the $j^{th}$ boson measured with respect to
the center of the trap.
We consider three different short-range model
potentials $V_{\rm{tb}}({\bf{r}}_{jk})$,
where ${\bf{r}}_{jk}={\bf{r}}_{j}-{\bf{r}}_{k}$.

Our two- and three-boson studies discussed in Secs.~\ref{sec_twobody}
and \ref{sec_threebody} employ
the regularized pseudopotential $V_{\rm{ps}}$~\cite{huan57},
\begin{eqnarray}
\label{eq_pseudoreg}
V_{\rm{ps}}({\bf r}_{jk})=\frac{4 \pi \hbar^2 a_s(k)}{m_a} 
\delta^{(3)}({\bf r}_{jk}) 
\frac{\partial}{\partial r_{jk}}r_{jk},
\end{eqnarray}
where $r_{jk}=|{\bf r}_{jk}|$.
The operator $(\partial/\partial r_{jk}) r_{jk}$ ensures that the 
$N$-particle wave function $\psi$
is well behaved when the interparticle distance $r_{jk}$ goes to zero.
In Eq.~(\ref{eq_pseudoreg}), $a_s(k)$ denotes the energy-dependent
scattering length~\cite{blum02,bold02},
\begin{eqnarray}
a_s(k) = -
\frac{\tan(\delta_0(k))}{k},
\end{eqnarray}
where $k$ denotes the wave vector
associated with the scattering energy $E_{\rm{sc}}^{\rm{rel}}$ in the relative
coordinate, $k=\sqrt{m_a E_{\rm{sc}}^{\rm{rel}}}/\hbar$,
and $\delta_0(k)$ the energy-dependent
$s$-wave scattering phase shift.
The ``usual'' (zero-energy) $s$-wave scattering length
is defined by taking the scattering energy to zero,
i.e., $a_s(0) =\lim_{k \rightarrow 0} a_s(k)$.
In many cases, 
the energy-dependence of $a_s(k)$ is 
neglegible and 
$a_s(k)$ can be
replaced by the zero-energy scattering length $a_s(0)$.
In other cases (see Secs.~\ref{sec_twobody} and \ref{sec_nboson}), 
it is convenient  to parameterize the
energy dependence of $a_s(k)$ in terms of the effective range $r_e$ 
and the shape or volume parameter $V$~\cite{newton},
\begin{eqnarray}
\label{eq_effrange}
\frac{1}{a_s(k)} = \frac{1}{a_s(0)} - \frac{1}{2}r_e k^2 +
\frac{1}{8}V k^4+ {\cal{O}}(k^6)
\end{eqnarray}
or
\begin{eqnarray}
\label{eq_effrange2}
{a_s(k)} = {a_s(0)} + \frac{1}{2} [a_s(0)]^2 r_e k^2 -
 \frac{1}{8} [a_s(0)]^2 V k^4+ {\cal{O}}(k^6).
\end{eqnarray}
We note that $a_s(0)$, $r_e$ and $V$ are only defined if the
two-body potential falls off faster than $r_{jk}^{-3}$,
$r_{jk}^{-5}$ and $r_{jk}^{-7}$, respectively, 
in the large $r_{jk}$ limit~\cite{mott65,levy63}.
The pseudopotential given in Eq.~(\ref{eq_pseudoreg}) 
can alternatively be parametrized through the boundary 
condition~\cite{wign33,beth35}
\begin{eqnarray}
\label{eq_pseudoregbc}
\left[
\frac{ \frac{\partial \left( r_{12} 
\psi({\bf r}_{12},{\bf R}_{12},{\bf r}_3,\cdots,{\bf r}_N) \right) }
{\partial r_{12}}}{r_{12} \psi({\bf r}_{12},{\bf R}_{12},{\bf r}_3,\cdots,{\bf r}_N) }
\right]_{r_{12} \rightarrow 0}
= -\frac{1}{a_s(k)},
\end{eqnarray}
where ${\bf R}_{12}=({\bf r}_1+{\bf r}_2)/2$.
The limit $r_{12} \rightarrow 0$ on the left
hand side of Eq.~(\ref{eq_pseudoregbc}) is taken while keeping the 
coordinates ${\bf R}_{12},{\bf r}_3,\cdots,{\bf r}_N$ fixed.
Analogous expressions hold for the other interparticle distances.

In our perturbative
calculations (see Sec.~\ref{sec_nboson}),
in contrast, we write $V_{\rm{tb}}$ as a sum of the unregularized
or bare
Fermi pseudopotential $V_{\rm{F}}$~\cite{ferm34},
\begin{eqnarray}
\label{eq_fermipseudo}
V_{\rm{F}}({\bf r}_{jk})=\frac{4 \pi \hbar^2 a_s(0)}{m_a} \delta^{(3)}({\bf r}_{jk}),
\end{eqnarray}
and the
zero-range potential $V'$~\cite{bean07,john12},
\begin{eqnarray}
\label{eq_pseudoreff}
V'({\bf{r}}_{jk})= -\frac{\pi \hbar^2 [a_s(0)]^2 r_e}{m_a} 
\left( 
\nabla_{{\bf{r}}_{jk}}^2 \delta^{(3)}({\bf{r}}_{jk}) +
\delta^{(3)}({\bf{r}}_{jk}) \nabla_{{\bf{r}}_{jk}}^2 
\right),
\end{eqnarray}
which accounts for the
effective 
range dependence.
The first and second Laplacian in Eq.~(\ref{eq_pseudoreff})
act to the left and right, respectively, and ensure that the
pseudopotential $V'$ is Hermitian.
While a pseudopotential that depends on the
shape parameter could be added,
it is not considered here since it leads to higher order
contributions in $a_{\rm{ho}}^{-n}$ than we are
considering in Sec.~\ref{sec_nboson}.
Since $V_{\rm{F}}$ and $V'$ are not regularized,
their use within  perturbation theory
 leads to divergencies, which can be cured
within the framework of renormalized perturbation theory
by introducing appropriate counterterms 
denoted by $W$.
We treat $V_F$ and $V'$ in second- and first-order perturbation theory
(see Appendix~\ref{appendixA}).
This implies that $W$ must contain a term proportional
to $[a_s(0)]^2$ that cures the divergencies arising 
from $V_F$; no divergencies in the energy arise when treating
$V'$ in first-order perturbation theory~\cite{john09,john12}.

Lastly, our numerical stochastic variational
calculations (see Sec.~\ref{sec_nboson}) 
employ a short-range Gaussian potential
$V_{\rm{g}}$ with range $r_0$ and depth $V_0$,
\begin{eqnarray}
\label{eq_gaussian}
V_{\rm{g}}(r_{jk}) = 
V_0 \exp \left[ -\left( \frac{r_{jk}}{\sqrt{2} r_0} \right)^2 \right].
\end{eqnarray}
For a fixed $r_0$, $V_0$ is adjusted so as to generate potentials
with different two-body $s$-wave scattering lengths $a_s(0)$.
Throughout, we limit ourselves to parameter combinations such that
$V_{\rm{g}}$ supports no two-body bound states in free space. This implies that
$V_{\rm{g}}$ is purely repulsive, i.e., $V_0>0$, for $a_s(0)>0$.
For $a_s(0)<0$, we have $V_0 < 0$.
The leading order and sub-leading order energy-dependence of $a_s(k)$,
parameterized by $r_e$ and $V$, respectively, depend on 
$r_0$.

\begin{figure}
\vspace*{+.9cm}
\includegraphics[angle=0,width=60mm]{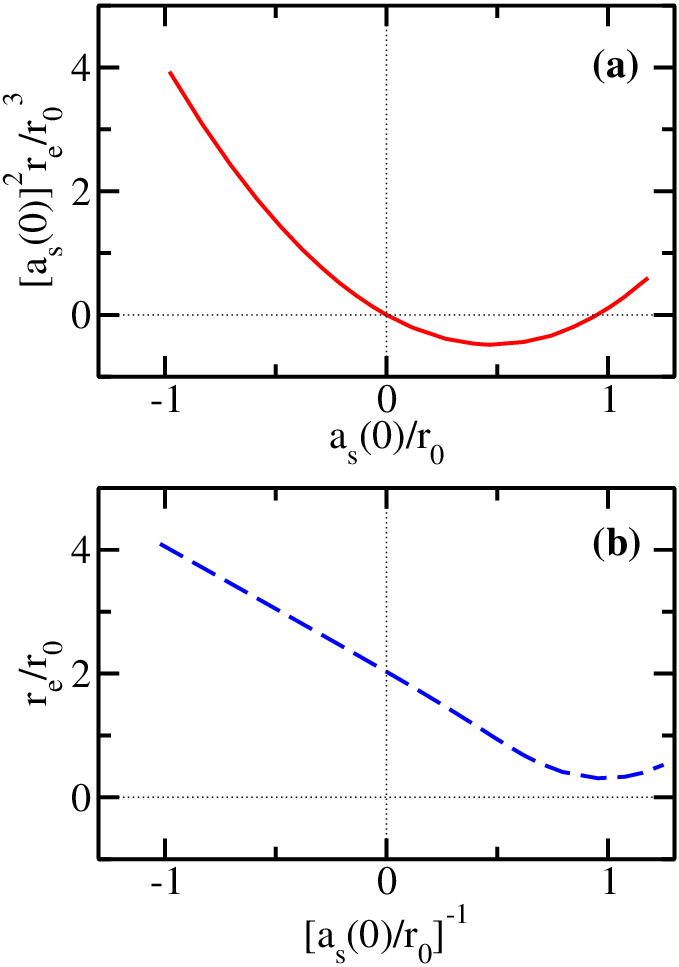}
\vspace*{0.2cm}
\caption{(Color online)
Scattering quantities for Gaussian model potential $V_{\rm{g}}$
in the ``weakly-interacting'' and ``strongly-interacting'' regimes.
(a) The solid line shows the quantity $[a_s(0)]^2r_e/r_0^3$
as a function of $a_s(0)/r_0$ (weakly-interacting regime).
(b) The dashed line shows the quantity $r_e/r_0$ as
a function of $[a_s(0)/r_0]^{-1}$ (strongly-interacting
regime).
Dotted lines are shown to enhance the readability of the graphs.
}\label{fig_gaussian}
\end{figure}

The solid line
in Fig.~\ref{fig_gaussian}(a) shows the quantity
$[a_s(0)]^2 r_e /r_0^3$ 
[see Eq.~(\ref{eq_effrange2})]
for the Gaussian model potential as a function
of the $s$-wave scattering length  $a_s(0)/r_0$.
It can be seen that
$[a_s(0)]^2 r_e /r_0^3$ goes to zero
as $|a_s(0)/r_0| \rightarrow 0$. Moreover, $[a_s(0)]^2 r_e /r_0^3$
is positive for small negative $a_s(0)/r_0$
and negative for small positive $a_s(0)/r_0$.
When $|a_s(0)|$ becomes infinitely large, the deviations from
universality depend on $r_e$ [see
Eq.~(\ref{eq_effrange})].
The dashed line in Fig.~\ref{fig_gaussian}(b) shows the effective range
$r_e /r_0$ 
for the Gaussian model potential as a function
of the inverse $s$-wave scattering length $[a_s(0)/r_0]^{-1}$.
The effective range $r_e/r_0$ 
is finite and positive
as 
$|a_s(0)| \rightarrow \infty$ and varies approximately linearly 
for small $[a_s(0)/r_0]^{-1}$ with negative 
slope.

The higher-order energy-dependence
of the $s$-wave scattering length
in the ``weakly-interacting''
and ``strongly-interacting'' regimes is
governed by the volume parameter $V$.
The quantity
$[a_s(0)]^2 V / r_0^5$ [see Eq.~(\ref{eq_effrange2})]
goes to zero as 
$|a_s(0)| \rightarrow 0$, and is positive 
for small $a_s(0)/r_0<0$ and negative for small $a_s(0)/r_0>0$.
The quantity $V/r_0^3$ [see Eq.~(\ref{eq_effrange})] 
is finite and negative when the 
$s$-wave scattering length
diverges;  $V/r_0^3$ varies approximately linearly 
for small $[a_s(0)/r_0]^{-1}$ with positive
slope.

Sections~\ref{sec_twobody}-\ref{sec_threebody}
present results for the occupation numbers $n_{{\boldsymbol{\nu}}}$, 
which are---for 
inhomogeneous systems---defined by way of the one-body density matrix 
$\rho({\bf r}_1',{\bf r}_1)$~\cite{penr56,lowd55,dubo01,blum11},
\begin{align}
\label{eq_onebody}
\rho( {\bf r}'_1, &{\bf r}_1 ) = N \times \\
&\frac
{
\int [ \psi( {\bf r}'_1, {\bf r}_2, \cdots, {\bf r}_N ) ]^{*}
\psi( {\bf r}_1, {\bf r}_2, \cdots, {\bf r}_N )
d^3 {\bf r}_2 \cdots d^3 {\bf r}_N
} 
{
\int |\psi( {\bf r}_1, \cdots, {\bf r}_N )|^2 
d^3 {\bf r}_1 \cdots d^3 {\bf r}_N 
}. \nonumber
\end{align}
The one-body density
matrix $\rho({\bf r}'_1,{\bf r}_1)$
can be expanded in terms of a complete orthonormal set,
\begin{eqnarray}
\label{eq_natorb}
\rho({\bf r}'_1,{\bf r}_1) =  \sum_{{\boldsymbol{\nu}}}
n_{{\boldsymbol{\nu}}} \phi^*_{{\boldsymbol{\nu}}}({\bf r}_1) 
\phi_{{\boldsymbol{\nu}}}({\bf r}'_1) ,
\end{eqnarray}
where ${\boldsymbol{\nu}}$ 
collectively denotes the three quantum numbers $\nu\lambda\mu$
needed to uniquely label the functions $\phi_{{\boldsymbol{\nu}}}({\bf r}_1)$
of the complete orthonormal set.
If we use spherical coordinates,
$\nu$ is the radial label, 
$\lambda$ 
the partial wave label and $\mu$ the corresponding projection
number.
The quantities 
$\phi_{{\boldsymbol{\nu}}}({\bf r}_1)$ and $n_{{\boldsymbol{\nu}}}$
are called natural orbitals and
occupation numbers, respectively.
Our normalization is chosen such that 
$\sum_{{\boldsymbol{\nu}}} n_{{\boldsymbol{\nu}}} = N$.
The largest occupation number $n_{\boldsymbol{\nu}}$
defines  the
condensate fraction $N_0/N$ of the $N$-boson system, i.e.,
$N_0/N = \mbox{max}(n_{{\boldsymbol{\nu}}}/N)$. 

In practice, it is convenient to define
partial wave projections $\rho_{\lambda\mu}(r'_1,r_1)$,
\begin{eqnarray}
\label{eq_projection}
\rho_{\lambda\mu}({r}'_1,{r}_1)= 
\int Y_{\lambda\mu}^*(\hat{r}'_1)
\rho({\bf r}'_1\!,{\bf r}_1) 
Y_{\lambda\mu}(\hat{r}_1) d^2\hat{r}'_1 d^2 \hat{r}_1,
\end{eqnarray}
where $d^2 \hat{r}_1$ and $d^2 \hat{r}'_1$
denote angular volume elements.
The
two-dimensional projections $\rho_{\lambda\mu}(r'_1,r_1)$ 
can be diagonalized,
yielding the occupation
numbers $n_{{\boldsymbol{\nu}}}=n_{\nu\lambda\mu}$ and 
the radial parts $P_{\nu\lambda}(r_1)$ of
the natural orbitals $\phi_{{\boldsymbol{\nu}}}({\bf r}_1)$, where
$P_{\nu\lambda}(r_1)$ is defined through
$\phi_{{\boldsymbol{\nu}}}({\bf r}_1)=P_{\nu\lambda}(r_1)Y_{\lambda\mu}(\hat{r}_1)$.

\section{Trapped two-body system}
\label{sec_twobody}
The eigen energies and eigen states of the two-particle Hamiltonian
are most easily determined by
transforming the Schr\"odinger equation 
to center of mass and relative 
coordinates
${\bf R}_{12}$ and ${\bf r}_{12}$.
In these coordinates, the wave function $\psi$ separates
into the center of mass wave function $\psi_{QLM}^{\rm{cm}}({\bf R}_{12})$
and
the relative wave function $\psi_{q lm}^{\rm{rel}}({\bf r}_{12})$. 
The two-body energy then reads
\begin{eqnarray}
E_2=E_2^{\rm{cm}}+E_2^{\rm{rel}},
\end{eqnarray}
where 
\begin{eqnarray}
E_2^{\rm{cm}}=(2Q+L+3/2)\hbar \omega
\end{eqnarray}
and
\begin{eqnarray}
\label{eq_defineq}
E_2^{\rm{rel}}=(2q+l+3/2) \hbar \omega
\end{eqnarray} 
with
$Q=0,1,\cdots$, $L=0,1,\cdots$ and
$l=0,1,\cdots$. For each $L$ ($l$), the 
center of mass (relative) energy has a $2L+1$ ($2l+1$)
degeneracy
that is associated with the projection quantum number $M$ ($m$).
The allowed values of $q$, 
and consequently the radial parts of the relative wave function
and 
the relative eigen energies, 
depend on the 
functional form of the interaction potential $V_{\rm{tb}}$.

\begin{figure}
\vspace*{+.9cm}
\includegraphics[angle=0,width=60mm]{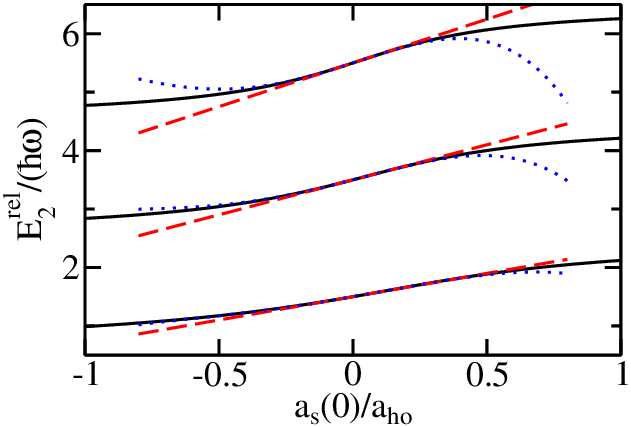}
\vspace*{0.2cm}
\caption{(Color online)
Solid lines show the
relative energies $E_2^{\rm{rel}}$ with $l=0$
for the trapped two-boson system
interacting through $V_{\rm{ps}}$,
Eq.~(\ref{eq_energyn2}) with $a_s(E_2^{\rm{rel}})$ replaced by $a_s(0)$,
as a function
of the $s$-wave scattering length $a_s(0)/a_{\rm{ho}}$.
The dashed and dotted lines show Eq.~(\ref{eq_NIexpansion}) 
for $i \le 1$ and $j=0$, and $i \le 4$ and $j=0$, respectively.
}\label{fig1}
\end{figure}

For the energy-dependent zero-range 
potential $V_{\rm{ps}}$ [see Eq.~(\ref{eq_pseudoreg})],
the relative wave functions with $l>0$ are not affected by
the interaction potential,
implying $q=0,1,\cdots$;
in this case, the relative wave function
coincides with the harmonic oscillator wave function
and the two-body energy is independent of 
the $s$-wave scattering length.
For $l=0$, the relative eigen energies
are obtained by solving the transcendental 
equation~\cite{busc98}
\begin{eqnarray}
\label{eq_energyn2}
\frac{\sqrt{2}\Gamma(3/4-E_2^{\rm{rel}}/(2 \hbar \omega))}
{\Gamma(1/4-E_2^{\rm{rel}}/(2 \hbar \omega))}
=\frac{a_{\rm{ho}}}
{a_s(E_2^{\rm{rel}})},
\end{eqnarray}
where the energy-dependent $s$-wave scattering length is evaluated at
the relative energy of the trapped system, i.e., where
we have set 
$E_{\rm{sc}}^{\rm{rel}}=E_2^{\rm{rel}}$~\cite{blum02,bold02}.
Solid lines in Fig.~\ref{fig1} show the relative eigen energies 
with $l=0$ obtained by solving Eq.~(\ref{eq_energyn2}) with
$a_s(E_2^{\rm{rel}})$ replaced by $a_s(0)$
as a function of
$a_s(0)/a_{\rm{ho}}$.
In general, the eigen energies of the trapped
two-body system need  to
be determined self-consistently 
since $E_2^{\rm{rel}}$ appears on the left and right hand sides
of Eq.~(\ref{eq_energyn2})~\cite{blum02,bold02}.

For $|a_s(E_2^{\rm{rel}})/a_{\rm{ho}}| \ll 1$, 
we Taylor expand Eq.~(\ref{eq_energyn2})
around 
the non-interacting relative energies $E_{2,n}^{\rm{ni}}$,
where $E_{2,n}^{{\rm ni}}=(2 n + 3/2) \hbar \omega$
with $n=0,1,\cdots$.
Replacing $1/a_s(E_2^{\rm{rel}})$ 
by the right hand side of Eq.~(\ref{eq_effrange}), 
we find
\begin{eqnarray}
\label{eq_NIexpansion}
E_2^{\rm{rel}} = E_{2,n}^{{\rm ni}} + 
\sum_{i=1,j=0,j<i}^{i+j \le 4}
d^{(i,j)}_{2,n} 
\left(\frac{a_s(0)}{a_{\rm{ho}}} \right)^i \!\! \left(\frac{r_e}{a_{\rm{ho}}} \right)^j
\hbar \omega \nonumber \\
+\cdots.
\end{eqnarray}
The next terms 
are
proportional to 
$[a_s(0)]^5$, $[a_s(0)]^4 r_e$, $[a_s(0)]^3 r_e^2$
and 
$[a_s(0)]^2 V$.
The coefficients $d^{(i,j)}_{2,n}$ 
can
be compactly written in terms of 
the quantity $h_{n,p}$,
\begin{eqnarray}
\label{eq_Hnp}
h_{n,p} = H_{n,p} + (-1)^p H_{-n-3/2,p},
\end{eqnarray}
where $H_{n,p}$ is a generalized harmonic number~\cite{notation1}.
Explicit expressions
for the
coefficients $d^{(i,j)}_{2,n}$ 
with $i+j \le 4$ are reported in Table~\ref{tab1}.
{\renewcommand{\arraystretch}{2}\!\!\!
\begin{table}
\caption{Expansion coefficients $d^{(i,j)}_{2,n}$, 
see Eq.~(\ref{eq_NIexpansion}), 
for the 
weakly-interacting
two-boson system with $s$-wave interactions.}
\begin{ruledtabular}
\begin{tabular}{ll|r}
$i$ & $j$ & $d^{(i,j)}_{2,n}$ \\
\hline
1 & 0 & $(-1)^{n+1} 2^{3/2}/\left[n! \; \Gamma\left(-n-1/2\right)\right]$ \\
2 & 0 & $-4 h_{n,1} /\left[n! \; \Gamma\left(-n-1/2\right)\right]^2$ \\
3 & 0 & $(-1)^{n+1} 2^{3/2} \left( h_{n,2} + 3 h_{n,1}^2 \right)/\left[n! \; \Gamma\left(-n-1/2\right)\right]^3$ \\
4 & 0 & $- \left(8/3\right) \left( h_{n,3} + 6 h_{n,2} h_{n,1} + 8 h_{n,1}^3 \right)/\left[n! \; \Gamma\left(-n-1/2\right)\right]^4$ \\ 
\hline
2 & 1 & $\frac{1}{2} (2n+3/2) d_{2,n}^{(1,0)}$ \\
3 & 1 & $\frac{1}{2} \left(d_{2,n}^{(1,0)} \right)^2 + (2n+3/2) d_{2,n}^{(2,0)}$ \\
\end{tabular}
\end{ruledtabular}
\label{tab1}
\end{table}}
The effective range enters first in combination with
the square of the zero-energy scattering length.
For the Gaussian potential $V_{\rm{g}}$ considered in
Sec.~\ref{sec_nboson}, the product $[a_s(0)]^2r_e$
goes to zero as $|a_s(0)| \rightarrow 0$ 
(see solid line in Fig.~\ref{fig_gaussian}).
Dashed and dotted lines in Fig.~\ref{fig1} show Eq.~(\ref{eq_NIexpansion})
with $j=0$ for $i \le 1 $ and $i \le 4$, respectively. 
The Taylor expanded expressions for the ground state with $i \le 1$
and $i \le 4$ agree  with the exact eigen energy to better than
1\% for
$-0.22 < a_s(0)/a_{\rm{ho}} < 0.54$
and $-0.67 < a_s(0)/a_{\rm{ho}}< 0.52$, respectively.
The accuracy of the Taylor
expansion deteriorates 
more quickly for the excited states.
References~\cite{john09,john12} discuss
the structure of Eq.~(\ref{eq_NIexpansion})
with $i+j \le 3$ for the ground state as well as extensions 
for $N>2$.

We also expand around the strongly-interacting regime.
For $|a_{\rm{ho}}/a_s(E_2^{\rm{rel}})| \ll 1$, 
we expand Eq.~\eqref{eq_energyn2} 
around 
the relative energies $E_{2,n}^{\rm{unit}}$ at unitarity,
where $E_{2,n}^{{\rm unit}}=(2 n + 1/2) \hbar \omega$
with $n=0,1,\cdots$.
Replacing $1/a_s(E_2^{\rm{rel}})$ 
by the right hand side of Eq.~(\ref{eq_effrange}), 
we find
\begin{eqnarray}
\label{eq_expansion_n2unit}
E_2^{\rm{rel}} = 
E_{2,n}^{{\rm unit}} + 
\sum_{i=0,j=0}^{i+j \le 3}
\tilde{d}^{(i,j)}_{2,n} 
\left(\frac{a_s(0)}{a_{\rm{ho}}} \right)^{\!\!\!-i} 
\!\!\! \left(\frac{r_e}{a_{\rm{ho}}} \right)^j \!\!\! \hbar \omega \nonumber \\
+ \frac{1}{8} \left(2n+\frac{1}{2}\right)^2 \tilde{d}^{(1,0)}_{2,n} \frac{V}{a_{\rm{ho}}^3} \hbar \omega 
+ \cdots.
\end{eqnarray}
Similarly to the weakly-interacting case, 
we find it convenient to express the expansion coefficients
$\tilde{d}_{2,n}^{(i,j)}$ in terms of the function 
\begin{eqnarray}
\tilde{h}_{n,p} = H_{n,p} + (-1)^p H_{-n-1/2,p}.
\end{eqnarray}
Table~\ref{tab2} shows the 
expansion coefficients $\tilde{d}_{2,n}^{(i,j)}$ with $i+j \le 3$.
{\renewcommand{\arraystretch}{2}\!\!\!\!\!
\begin{table}
\caption{Expansion coefficients $\tilde{d}^{(i,j)}_{2,n}$, 
see Eq.~(\ref{eq_expansion_n2unit}), 
for the 
strongly-interacting
two-boson system with $s$-wave interactions.}
\begin{ruledtabular}
\begin{tabular}{ll|r}
$i$ & $j$ & $\tilde{d}^{(i,j)}_{2,n}$ \\ 
\hline
0 & 0 & 0 \\
1 & 0 & $(-1)^{n+1} 2^{3/2}/\left[2\;n! \; \Gamma\left(-n+1/2\right)\right]$ \\
2 & 0 & $-4 \tilde{h}_{n,1} /\left[2\;n! \; \Gamma\left(-n+1/2\right)\right]^2$ \\
3 & 0 & $(-1)^{n+1} 2^{3/2} \left( \tilde{h}_{n,2} + 3 \tilde{h}_{n,1}^2 \right)/\left[2\;n! \; \Gamma\left(-n+1/2\right)\right]^3$ \\
\hline
0 & 1 & $-\frac{1}{2}(2n+1/2)\tilde{d}_{2,n}^{(1,0)} $ \\
1 & 1 & $- \left[\frac{1}{2} \left(\tilde{d}_{2,n}^{(1,0)}\right)^2 + (2n+1/2) \tilde{d}_{2,n}^{(2,0)}\right]$ \\
2 & 1 & $-\frac{3}{2} \left[\tilde{d}_{2,n}^{(1,0)} \tilde{d}_{2,n}^{(2,0)} + (2n+1/2)\tilde{d}_{2,n}^{(3,0)}\right]$ \\
\hline
0 & 2 & $\frac{1}{4} (2n+1/2) \left[\left(\tilde{d}_{2,n}^{(1,0)}\right)^2 + (2n+1/2) \tilde{d}_{2,n}^{(2,0)}\right]$ \\
1 & 2 & $\frac{1}{4} \bigg[\left(\tilde{d}_{2,n}^{(1,0)}\right)^3 + 6 (2n+1/2) \tilde{d}_{2,n}^{(1,0)} \tilde{d}_{2,n}^{(2,0)} \qquad$ \\
  &   & $+ 3 \left(2n+1/2\right)^2 \tilde{d}_{2,n}^{(3,0)}\bigg]$ \\
\hline
0 & 3 & $-\frac{1}{8} (2n+1/2) \bigg[\left(\tilde{d}_{2,n}^{(1,0)}\right)^3 + 3 (2n+1/2) \tilde{d}_{2,n}^{(1,0)} \tilde{d}_{2,n}^{(2,0)} \qquad$ \\
  &   & $+ \left(2n+1/2\right)^2 \tilde{d}_{2,n}^{(3,0)}\bigg]$ \\
\end{tabular}
\end{ruledtabular}
\label{tab2}
\end{table}}
Equation~(\ref{eq_expansion_n2unit}) 
contains contributions
that are directly proportional to the inverse scattering length $1/a_s(0)$,
the effective range $r_e$ and the volume term. Which of these
terms dominates depends on the interaction potential.
Importantly, while the effective range diverges for 
the Gaussian model potential
in the $|a_s(0)| \rightarrow 0$ limit, it remains 
finite in the $|a_s(0)|^{-1} \rightarrow 0$ limit
[see dashed line in Fig.~\ref{fig_gaussian}(b)].
The next order terms in Eq.~(\ref{eq_expansion_n2unit}) are
proportional to
$[a_s(0)]^{-i} r_e^{j} V^k$ with $i+j+3k=4$.

Next we discuss the occupation numbers of the two-boson system.
The determination of the one-body density matrix 
for the two-boson system requires
that the wave function $\psi$,
written above in terms of the center of mass and relative coordinates
${\bf R}_{12}$ and ${\bf r}_{12}$, 
be transformed to 
the single particle coordinates ${\bf r}_1$ and ${\bf r}_2$.
In the following, we discuss
the occupation numbers associated with 
the one-body density matrix as a function
of the relative two-body energy $E_2^{\rm{rel}}$,
assuming that the center of mass wave function
is in the ground state, i.e., we set $Q=L=M=0$.
For two-boson systems in one dimension, the one-body density matrix 
was calculated as a function of 
temperature in Ref.~\cite{ciro01}.
Here, we consider the zero temperature limit
and restrict
ourselves to relative states with $l=0$.
The formalism developed, however, can be straighforwardly applied to
states with finite $l$, $Q$, $L$ and $M$.
As detailed in Appendix~\ref{appendixB},
the one-body density matrix can be evaluated
efficiently and with high accuracy by expanding it in terms of 
single particle 
harmonic oscillator functions.

\begin{figure}
\vspace*{+.9cm}
\includegraphics[angle=0,width=60mm]{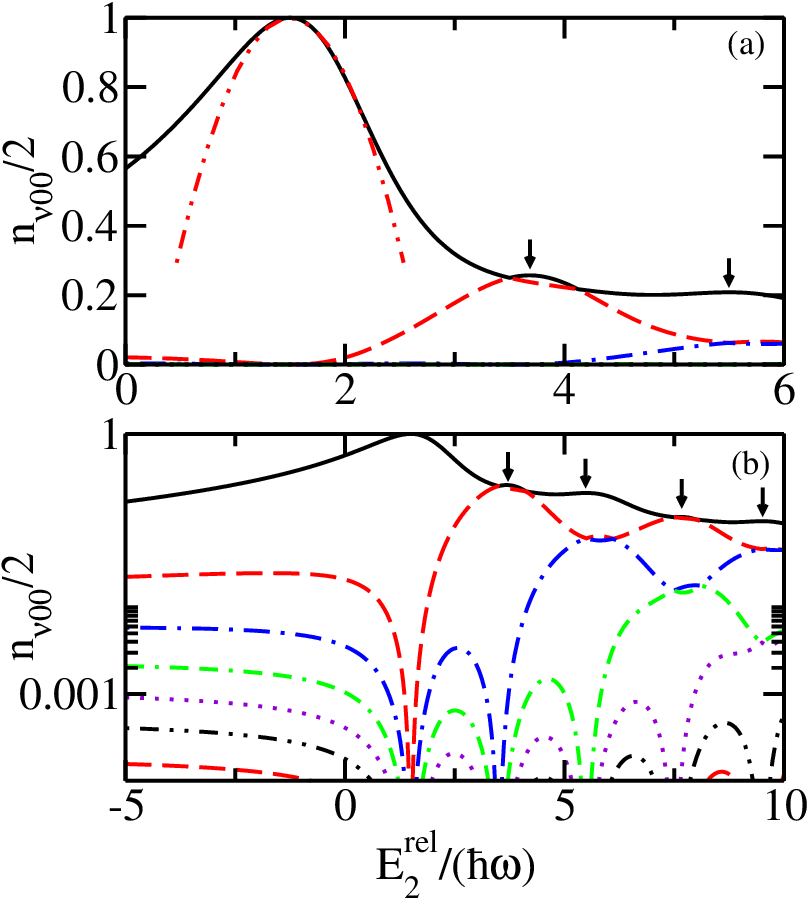}
\vspace*{0.2cm}
\caption{(Color online)
Occupation numbers per particle $n_{\nu 00}/N$ of the two-boson system
interacting through
$V_{\rm{ps}}$ with $a_s(E_2^{\rm{rel}})$ 
replaced by $a_s(0)$ 
as a function of the relative two-body energy $E_2^{\rm{rel}}$ for $Q=L=M=0$.
(a) Solid, dashed and dash-dotted lines show
$n_{\nu 00}/2$ for $\nu=0,1$ and $2$, respectively, on
a linear scale.
The dash-dot-dotted line shows the 
leading-order depletion,
i.e., the first two terms on the right hand side of
Eq.~(\ref{eq_condfracn2}), 
near $E_2^{\rm{rel}}=3 \hbar\omega/2$.
(b)
Lines show 
$n_{\nu 00}/2$ for $\nu=0-6$, from top to bottom
at $E_2^{\rm{rel}}=-5 \hbar \omega$,
on
a log scale.
Arrows mark the local maxima of $n_{\nu00}/2$.
Near the non-interacting energies, 
$(E_2^{\rm{rel}}-E_{2,n}^{\rm{rel}})/(\hbar \omega)$ is
to a good approximation directly proportional to the
$s$-wave scattering length $a_s(0)/a_{\rm{ho}}$ [see
Eq.~(\ref{eq_NIexpansion})], implying that the figure
can be read as a ``scaled occupation number versus scattering length''
plot.
}\label{fig_occupationn2}
\end{figure}

We first consider the zero-range pseudopotential $V_{\rm{ps}}$ 
with $a_s(E_2^{\rm{rel}})$ 
replaced by $a_s(0)$.
Figure~\ref{fig_occupationn2}
shows the scaled occupation numbers $n_{\nu00}/2$ 
for the trapped two-boson system interacting through
$V_{\rm{ps}}$ 
as a function
of the relative two-body energy $E_2^{\rm{rel}}$.
In the non-interacting limit, 
the ground 
state with energy $E_2^{\rm{rel}}=3 \hbar\omega /2$
is characterized by a single natural orbital with projection $\lambda=0$,
i.e., the non-interacting two-boson system in the ground state
has a condensate fraction $N_0/N$ of 1.
As the interactions are turned on, i.e.,
as $a_s(0)$ takes on small positive or negative
values corresponding to 
$E_2^{\rm{rel}}>3 \hbar\omega /2$ and
$E_2^{\rm{rel}}<3 \hbar\omega /2$, respectively, 
the occupation number associated with the lowest
natural orbital depletes. Taylor-expanding the
projected one-body density matrix around
$E_2^{\rm{rel}}=3 \hbar\omega /2$
(see 
Appendix~\ref{appendixB}),
we find that the condensate fraction of the ground state 
of the two-boson system depletes 
quadratically with %$E_2^{\rm{rel}}$, or equivalently 
$a_s(0)$, 
\begin{align}
\label{eq_condfracn2}
N_0/2 = 1 & - 0.420004 [a_s(0)/a_{\rm{ho}}]^2  \nonumber \\
& -0.373241 [a_s(0)/a_{\rm{ho}}]^3 \nonumber \\
& +0.406786 [a_s(0)/a_{\rm{ho}}]^4 \nonumber \\
& + {\cal{O}}([a_s(0)/a_{\rm{ho}}]^5).
\end{align}
The dash-dot-dotted line in 
Fig.~\ref{fig_occupationn2}(a)
shows the first two terms of Eq.~(\ref{eq_condfracn2}),
i.e., the leading-order dependence of the depletion on $a_s(0)$,
in the weakly-interacting regime.
The higher-order corrections proportional
to $[a_s(0)]^i$, $i=3$ and 4,  are analyzed
in Sec.~\ref{sec_nboson}.

Figure~\ref{fig_occupationn2} reveals oscillatory behavior
of the scaled occupation numbers $n_{\nu 00}/2$
for $E_2^{\rm{rel}} > 3 \hbar \omega/2$.
As the relative energy $E_2^{\rm{rel}}$
($E_2^{\rm{rel}}>3 \hbar \omega/2$) increases, the occupation numbers
go through ``near deaths and revivals'', 
with many of the occupation numbers
crossing.
When a higher-lying non-interacting state is reached, one more natural
orbital with $\lambda=0$ becomes macroscopically occupied.
Similar structure is seen for $\lambda >0$ (not shown).
For
$E_2^{\rm{rel}}=7 \hbar\omega/2$, 
e.g., five natural orbitals are occupied.
Two of these natural orbitals have $\lambda=0$ with 
$n_{\nu 00}/2=1/4$ 
(solid and dashed lines in Fig.~\ref{fig_occupationn2}),
while three natural orbitals 
(from the $2 \lambda +1$ degeneracy)
have $\lambda=1$ with $n_{\nu 00}/2 = 1/6$.
The largest occupation number per particle $n_{000}/2$ 
takes on local maxima
at the non-interacting energies 
$E_2^{\rm{rel}}=11\hbar \omega/2$ and $19\hbar \omega/2$
as well as at
$E_2^{\rm{rel}} \approx 3.68 \hbar \omega$ and $7.64 \hbar \omega$
(see arrows in Fig.~\ref{fig_occupationn2}).

For $E_2^{\rm{rel}} < 3 \hbar \omega/2$, the scaled
occupation number 
$n_{000}/2$ 
(see solid line
in Fig.~\ref{fig_occupationn2})
decreases monotonically with decreasing energy while 
many other natural orbitals become occupied, including natural
orbitals with $\lambda >0$. In the limit of a deeply bound two-body
state, the relative two-body wave function becomes infinitely
sharply peaked, implying that infinitely many single-particle states
are required to describe the deeply-bound two-boson system.

If the energy-dependence of the $s$-wave scattering length is 
accounted for,
the condensate fraction of the two-boson system in the
ground state near the non-interacting regime
depends not only on $a_s(0)$, see Eq.~(\ref{eq_condfracn2}),
but also on $r_e$. 
We find that
the leading-order effective range contribution
to the condensate fraction is 
$-(3/2) \times 0.420004 r_e[a_s(0)]^3/a_{\rm{ho}}^4$.
The factor of $3/2$ arises since the scattering length
$a_s(0)$ has to be replaced, following Eq.~(\ref{eq_effrange2}),
by $[a_s(0)]^2 r_e k^2/2$, where the relevant energy scale for evaluating
$k^2$ is $3 \hbar \omega/2$ [see Eq.~(\ref{eq_NIexpansion})
for $n=0$].

\section{Weakly-interacting trapped $N$-boson gas}
\label{sec_nboson}
Subsection~\ref{sec_nboson_a} determines the condensate
fraction of the lowest gas-like state of the $N$-boson system
perturbatively in
the weakly-interacting regime,
and
compares the perturbative
predictions with our
results for finite-range potentials.
Subsection~\ref{sec_nboson_b}  parametrizes
and quantifies the non-universal
corrections revealed through the comparison.

\subsection{Perturbative treatment}
\label{sec_nboson_a}
We employ the formalism of second quantization
and rewrite 
the $N$-boson Hamiltonian
as
\begin{eqnarray}
\label{eq_hamsecond}
H = 
\sum_{{\bf a}} 
E_{{\bf a}} \hat{a}^{\dagger}_{{\bf a}} 
\hat{a}_{{\bf a}} + \frac{1}{2} 
\sum_{{\bf a} {\bf b} {\bf c} {\bf d}}
K_{{\bf a} {\bf b} {\bf c} {\bf d}}
\hat{a}^{\dagger}_{{\bf a}} \hat{a}^{\dagger}_{{\bf b}}
\hat{a}_{{\bf d}}  \hat{a}_{{\bf c}},
\end{eqnarray}
where
\begin{align}
\label{eq_kmatrix}
 K_{{{\bf a}} {{\bf b}}} & \;\!\!_{{{\bf c}} {{\bf d}}} = 
\\
& \int \!\!\!\!\! 
\int \!\! \Phi_{{{\bf a}}}^*({\bf r}_1) \Phi_{{{\bf b}}}^*({\bf r}_2)
V_{\rm{tb}}({\bf r}_1-{\bf r}_2) 
\Phi_{{{\bf c}}}({\bf r}_1) \Phi_{{{\bf d}}}({\bf r}_2) 
d^3{\bf r}_1 d^3{\bf r}_2.\nonumber
\end{align}
Here, the $\Phi_{{{\bf a}}}({\bf r})$ denote the single particle
harmonic oscillator wave functions with 
eigen energy $E_{\bf{a}}$;
in spherical coordinates, we have 
$E_{{\bf a}} = (2 n_a+l_a+3/2)\hbar \omega$.
The operators $\hat{a}_{{\bf a}}$ and $\hat{a}_{{\bf a}}^{\dagger}$
obey bosonic commutation relations and
respectively annihilate and create a boson in the
single particle state $\Phi_{{\bf{a}}}$.
We model the interaction $V_{\rm{tb}}$
by the sum $V_{\rm{F}}+V'$ (see Sec.~\ref{sec_system}),
and employ the counterterms derived in
Refs.~\cite{john09,john12} to cure divergencies. 
Since the $\Phi_{{\bf a}}({\bf r})$ are known,
the matrix elements $K_{{\bf a} {\bf b} {\bf c} {\bf d}}$ 
for this interaction model
can be 
evaluated either analytically or numerically~\cite{john09,john12}.
The low-energy Hamiltonian given in Eq.~(\ref{eq_hamsecond})
has previously been used to derive perturbative energy
expressions up
to order $a_{\rm{ho}}^{-3}$ for the 
harmonically trapped $N$-boson system~\cite{john12}. 
In particular, $V_{\rm{F}}$ and $V'$ were treated at the level
of
third- and first-order perturbation theory.
Comparison with
energies for systems with finite-range interactions 
validated the formalism and
showed that the derived perturbative energy expressions, which depend
on $a_s(0)$ and $r_e$, provide an excellent description in the 
weakly-interacting regime~\cite{john12}.

To determine the condensate fraction, we construct 
the matrix $\langle \hat{a}_{\bf p}^{\dagger} \hat{a}_{\bf q} \rangle$,
where ${\bf p}$ and ${\bf q}$ 
run over all possible single-particle state labels.
The expectation value  
$\langle \hat{a}_{\bf p}^{\dagger} \hat{a}_{\bf q} \rangle$,
\begin{eqnarray}
\label{eq_ptwavefunction_matrix}
\langle \hat{a}_{\bf p}^{\dagger} \hat{a}_{\bf q} \rangle = 
\frac{\langle \psi_{\bf{0}}^{(k)} |
\hat{a}_{\bf p}^{\dagger} \hat{a}_{\bf q} | 
\psi_{\bf{0}}^{(k)} \rangle}
{\langle \psi_{\bf{0}}^{(k)} |  \psi_{\bf{0}}^{(k)} \rangle},
\end{eqnarray}
is
calculated with respect to the
many-body ground state wave function
$\psi_{\bf{0}}^{(k)}$,
determined within $k$$^{\rm{th}}$-order perturbation theory.
The ground state wave function
$\psi_{\bf{0}}^{(k)}$ is expressed as a superposition of the
unperturbed many-body wave functions $\psi_{\bf{j}}^{(0)}$, 
\begin{eqnarray}
\label{eq_ptwavefunction}
|\psi_{\bf{0}}^{(k)} \rangle = 
\sum_{\bf j} b_{\bf j}^{(k)} | \psi_{\bf j}^{(0)} \rangle,
\end{eqnarray}
where the expansion coefficients 
$b_{\bf j}^{(k)}$ are determined by
the matrix elements $K_{{\bf{a}}{\bf{b}}{\bf{c}}{\bf{d}}}$.
The subscript ${\bf j}$ collectively
labels the non-interacting or unperturbed many-body states
(in particular, ${\bf{0}}$ labels the ground state).
Once the matrix 
$\langle \hat{a}_{\bf{p}}^{\dagger} \hat{a}_{\bf{q}} \rangle$
is constructed, we diagonalize it
analytically (see Appendix~\ref{appendixA}). 
Up to order $a_{\rm{ho}}^{-3}$, i.e., treating the 
second-order
perturbation theory wave function, we find
\begin{align}
\label{eq_condpt}
N_0/N =1 & - 0.420004 (N-1)
\left[ \frac{a_s(0)}{a_{\rm{ho}}}\right]^2
\nonumber \\
& + \big[-0.373241
(N-1) \nonumber \\
& \quad\;\; + 0.439464
(N-1)(N-2)
\big] \left[\frac{a_s(0)}{a_{\rm{ho}}}\right]^3 \nonumber \\
& + \cdots .
\end{align}
The prefactors are discussed in the
context of
Table~\ref{tab_coefficient}.
We interpret the terms 
proportional to
$(N-1)$ and $(N-1)(N-2)$ as being due to two-body
and three-body scattering processes, respectively.
In Eq.~(\ref{eq_condpt}),
the terms proportional to $[a_s(0)]^2$ and $[a_s(0)]^3$
arise 
when treating the potential $V_{\rm{F}}$, together with the appropriate 
counterterm,
in first- and second-order perturbation theory.
Treating $V_{\rm{F}}$ 
in third-order perturbation theory (not pursued here),
three terms 
proportional to $[a_s(0)]^4$ that contain
the factors $(N-1)$, $(N-1)(N-2)$ and $(N-1)(N-2)(N-3)$, respectively,
are expected to arise.

The potential $V'$ does not,
in first-order perturbation theory, 
give rise to a two-body term proportional to
$r_e [a_s(0)]^2/a_{\rm{HO}}^3$.
This result agrees with that obtained by 
Taylor-expanding the
full two-body density matrix and determining its largest eigen value
(see last paragraph of Sec.~\ref{sec_twobody}).
No three-body term arises at order $r_e[a_s(0)]^3$.
Since the leading-order effective range
dependence is of order $a_{\rm{ho}}^{-4}$, it is
not included in Eq.~(\ref{eq_condpt}).

To assess the applicability of 
our perturbatively derived result,
Eq.~(\ref{eq_condpt}),
we calculate the condensate fraction for small
$N$-boson systems interacting through the Gaussian model 
potential $V_{\rm{g}}$,
Eq.~(\ref{eq_gaussian}), with small $|a_s(0)|/a_{\rm{ho}}$.
For $N=2$, we solve the relative Schr\"odinger equation using
standard B-spline techniques. For $N=3$ and $4$,
we use the stochastic variational approach~\cite{cgbook,sore05},
which expands the relative
eigen functions in terms of a set of fully symmetrized basis functions
whose widths are chosen semi-stochastically. 
The widths are optimized by minimizing the
ground state energy.
The optimized ground state wave function is then used
to construct the projected one-body density matrix $\rho_{00}(r'_1,r_1)$
on a grid in the $r'_1$ and $r_1$ coordinates.
The
projected one-body
density matrix
is diagonalized numerically
to find the natural orbitals and their occupation numbers.
The resulting condensate fraction for $N=3$
has an estimated numerical error of order $10^{-7}$ or smaller
for the parameter combinations considered. The numerical error
is due to the facts that 
(i) we use a finite basis set, 
(ii) we construct $\rho_{00}(r'_1,r_1)$ on a grid with finite
grid spacings, and 
(iii) our grid terminates at finite $r'_1$ and $r_1$ values.
For $N=3$, we use around 120 basis functions
and 625 linearly spaced grid
points in $r'_1$ and $r_1$ between $r_0/2$ and $5.5a_{\rm{ho}}$.

\begin{figure}
\vspace*{+.9cm}
\includegraphics[angle=0,width=80mm]{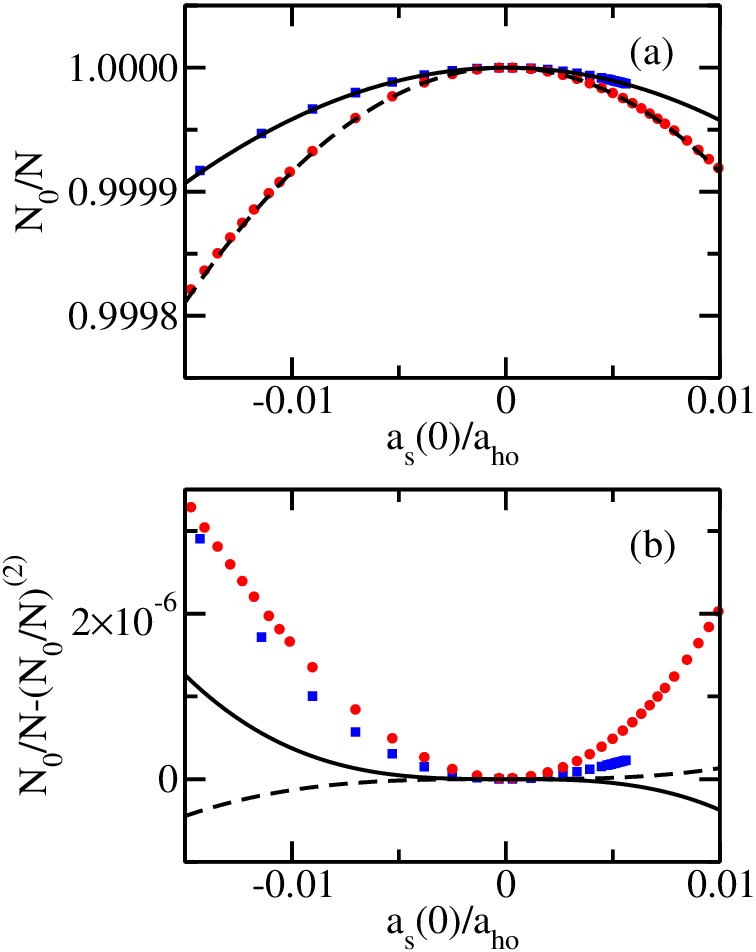}
\vspace*{0.2cm}
\caption{(Color online)
Condensate fraction $N_0/N$ of the weakly-interacting 
$N$-boson system as a function of $a_s(0)/a_{\rm{ho}}$.
(a)
Squares and circles show the condensate
fraction for $N=2$ and $3$ bosons
interacting through the shape-dependent Gaussian model
potential $V_{\rm{g}}$
with $r_0=0.01a_{\rm{ho}}$. Solid and dashed lines show
Eq.~(\ref{eq_condpt}) for $N=2$ and $3$, respectively.
(b)
Squares and circles show the
quantity $N_0/N-(N_0/N)^{(2)}$
for $N=2$ and 3 using the same data as in panel~(a).
Solid and dashed lines show 
the $[a_s(0)/a_{\rm{ho}}]^3$ term of
Eq.~(\ref{eq_condpt}) for $N=2$ and $3$.
}\label{fig_condweak}
\end{figure}

Squares and circles in Fig.~\ref{fig_condweak}(a)
show the condensate fraction $N_0/N$ for the $N=2$ and $N=3$ 
systems interacting through $V_{\rm{g}}$ 
with $r_0=0.01a_{\rm{ho}}$ as a function of the 
zero-energy $s$-wave scattering length $a_s(0)$.
For comparison, solid and dashed lines show our perturbative results 
up to order $a_{\rm{ho}}^{-3}$.
It can be seen that the numerically determined condensate depletions
for the finite-range interaction potential 
change approximatically quadratically with
the scattering length.
To investigate the correction
proportional to
$[a_s(0)/a_{\rm{ho}}]^3$,
squares and circles in Fig.~\ref{fig_condweak}(b) show the 
quantity $N_0/N-(N_0/N)^{(2)}$,
where $(N_0/N)^{(2)}=1-(N-1)0.420004  [a_s(0)/a_{\rm{ho}}]^2$, for
$N=2$ and 3 [the data for $N_0/N$ are the same as those shown
in Fig.~\ref{fig_condweak}(a)]. 
For comparison, the solid and dashed lines show the perturbatively
predicted 
$[a_s(0)/a_{\rm{ho}}]^3$ dependence for $N=2$ and $3$. 
It can be seen that the perturbative prediction does not
provide a good description of the sub-leading 
dependence of the depletion. 
Similar behavior is observed for $N=4$ (not shown).
We find that 
neither the inclusion of the $r_e[a_s(0)]^3$ nor of the $[a_s(0)]^4$
terms can explain the discrepancy.
As shown in the next subsection, the discrepancy
displayed in Fig.~\ref{fig_condweak}(b) is due
to non-universal contributions.

\subsection{Non-universal contributions}
\label{sec_nboson_b}
To better connect the results obtained
by applying perturbation theory to the low-energy
Hamiltonian, Eq.~(\ref{eq_hamsecond})
with $V_{\rm{tb}}=V_{\rm{F}}+V'$,
with the results for the Hamiltonian with finite-range
interactions, we first consider the two-body system.
Noting that the perturbative results for the condensate fraction
agree with the results obtained by Taylor-expanding the 
exact one-body density matrix for the two-body
system with regularized zero-range interaction
[i.e., noting that Eq.~(\ref{eq_condpt}) agrees with
Eq.~(\ref{eq_condfracn2}) for $N=2$ and the orders considered],
we compare the one-body density matrix
$\rho({\bf{r}}'_1,{\bf{r}}_1)$ for the two-body
system interacting through $V_{\rm{ps}}$ with
the one-body density matrix
$\rho_{\rm{fr}}({\bf{r}}'_1,{\bf{r}}_1)$ 
for the two-body system interacting through
a finite-range potential.

Since the
regularized zero-range potential reproduces the
relative two-body energy of systems with finite-range interactions
with high accuracy~\cite{bold02,blum02}, we
assume that the relative two-body energies
$E_2^{\rm{rel}}$ agree for the two interaction models.
We denote the normalized relative wave function
of the energetically lowest-lying gas-like state
for the finite-range potential by $\psi^{\rm{rel,fr}}({\bf{r}}_{12})$
and that for the regularized zero-range model
potential by $\psi_{q00}^{\rm{rel}}({\bf{r}}_{12})$ [see
Eq.~(\ref{eq_psi_r12INT})].
It is instructive to
write $\psi^{\rm{rel,fr}}$
as
\begin{eqnarray}
\label{eq_psig}
\psi^{\rm{rel,fr}}({\bf{r}}_{12}) = 
\psi_{q00}^{\rm{rel}}({\bf{r}}_{12}) + \delta \psi({\bf{r}}_{12}).
\end{eqnarray}
Inserting Eq.~(\ref{eq_psig}) into
$\rho_{\rm{fr}}({\bf{r}}'_1,{\bf{r}}_1)$ and assuming the
absence of center of mass excitations,
we find
\begin{eqnarray}
\label{eq_onebodyg}
\rho_{\rm{fr}}({\bf{r}}'_1,{\bf{r}}_1) \approx
\rho({\bf{r}}'_1,{\bf{r}}_1)+
2 \delta \rho({\bf{r}}'_1,{\bf{r}}_1),
\end{eqnarray}
where
$\rho({\bf{r}}'_1,{\bf{r}}_1)$ is given in 
Eqs.~(\ref{eq_step1}) and (\ref{eq_OBDM}),
and
\begin{align}
\label{eq_correction1}
\delta \rho({\bf{r}}'_1&, {\bf{r}}_1)=\\
&\int [
\psi^{\rm{cm}}_{000}({\bf{R}}'_{12}) 
\delta \psi({\bf{r}}'_{12}) 
]^*
\psi^{\rm{cm}}_{000}({\bf{R}}_{12}) 
\psi_{q00}^{\rm{rel}}({\bf{r}}_{12}) 
d^3 {\bf{r}}_2
\nonumber \\
&+\int [
\psi^{\rm{cm}}_{000}({\bf{R}}'_{12}) 
\psi_{q00}^{\rm{rel}}({\bf{r}}'_{12}) 
]^*
\psi^{\rm{cm}}_{000}({\bf{R}}_{12}) 
\delta \psi({\bf{r}}_{12}) 
d^3 {\bf{r}}_2 \nonumber
\end{align}
with
${\bf{r}}'_{12}={\bf{r}}_1'-{\bf{r}}_2$ and 
${\bf{R}}'_{12}=({\bf{r}}_1'+{\bf{r}}_2)/2$;
${\bf{r}}_{12}$ and ${\bf{R}}_{12}$ are defined in Sec.~\ref{sec_system}.
In writing Eq.~(\ref{eq_onebodyg}),
the term proportional to 
$|\delta \psi|^2$ has been neglected.

To determine the condensate fraction, we expand
$\delta \rho({\bf{r}}'_1,{\bf{r}}_1)$
in terms of non-interacting single particle harmonic oscillator functions;
this approach is analogous to that discussed in detail in
Appendix~\ref{appendixB} for $\rho({\bf{r}}'_1,{\bf{r}}_1)$.
A fairly straightforward analysis shows that 
the main correction to the condensate fraction arises
from the
%$(000,000)$ 
$((0,0,0),(0,0,0))$ 
element of $\delta \rho({\bf{r}}'_1,{\bf{r}}_1)$.
It follows that the largest occupation
number $n_{000}^{\rm{fr}}$ for the
finite-range potential can be written as
\begin{eqnarray}
\label{eq_condn2all}
\frac{n_{000}^{\rm{fr}}}{2}\approx
\frac{n_{000}}{2}+
\delta c_{00},
%^{0},
\end{eqnarray}
where
\begin{eqnarray}
\label{eq_correctioncond}
\delta c_{00}
%^0
= \sum_{i=0}^{\infty} 2^{-2i} (C_i^* D_i+C_i D_i^*).
\end{eqnarray}
In Eq.~(\ref{eq_condn2all}),
$n_{000}/2=N_0/2$ denotes the condensate fraction of the two-body system
interacting through the regularized zero-range potential
[see Eq.~(\ref{eq_condfracn2})].
The coefficients $C_i$ are defined in Eq.~(\ref{eq_coeffpseudo})
and the $D_i$ denote the overlaps between the non-interacting harmonic
oscillator functions and $\delta \psi$,
\begin{eqnarray}
\label{eq_u0trap}
D_i = 
\int [\psi_{i00}^{\rm{rel,ni}}({\bf{r}}_{12})]^* 
\delta \psi({\bf{r}}_{12})
d^3 {\bf{r}}_{12}.
\end{eqnarray}
Realizing that the $i=0$ terms in Eq.~(\ref{eq_correctioncond})
dominate and using the leading-order behavior of $C_0$, i.e.,
$C_0\approx 1$ [see Eqs.~(\ref{eq_Cq_i0_fourthOrder}) 
and~\eqref{eq_Cq_coeff1}],
we find
\begin{eqnarray}
\label{eq_correctioncond2}
\delta c_{00}
%^0
\approx
 2 {\mbox{Re}}(D_0).
\end{eqnarray}
For the finite-range potentials considered in this
paper, we find that Eqs.~(\ref{eq_correctioncond}) and
(\ref{eq_correctioncond2}) deviate by less than 
0.2~\%.
In the following, we refer to $D_0$ as the 
``non-universal two-body parameter''.

\begin{figure}
\vspace*{+.9cm}
\includegraphics[angle=0,width=80mm]{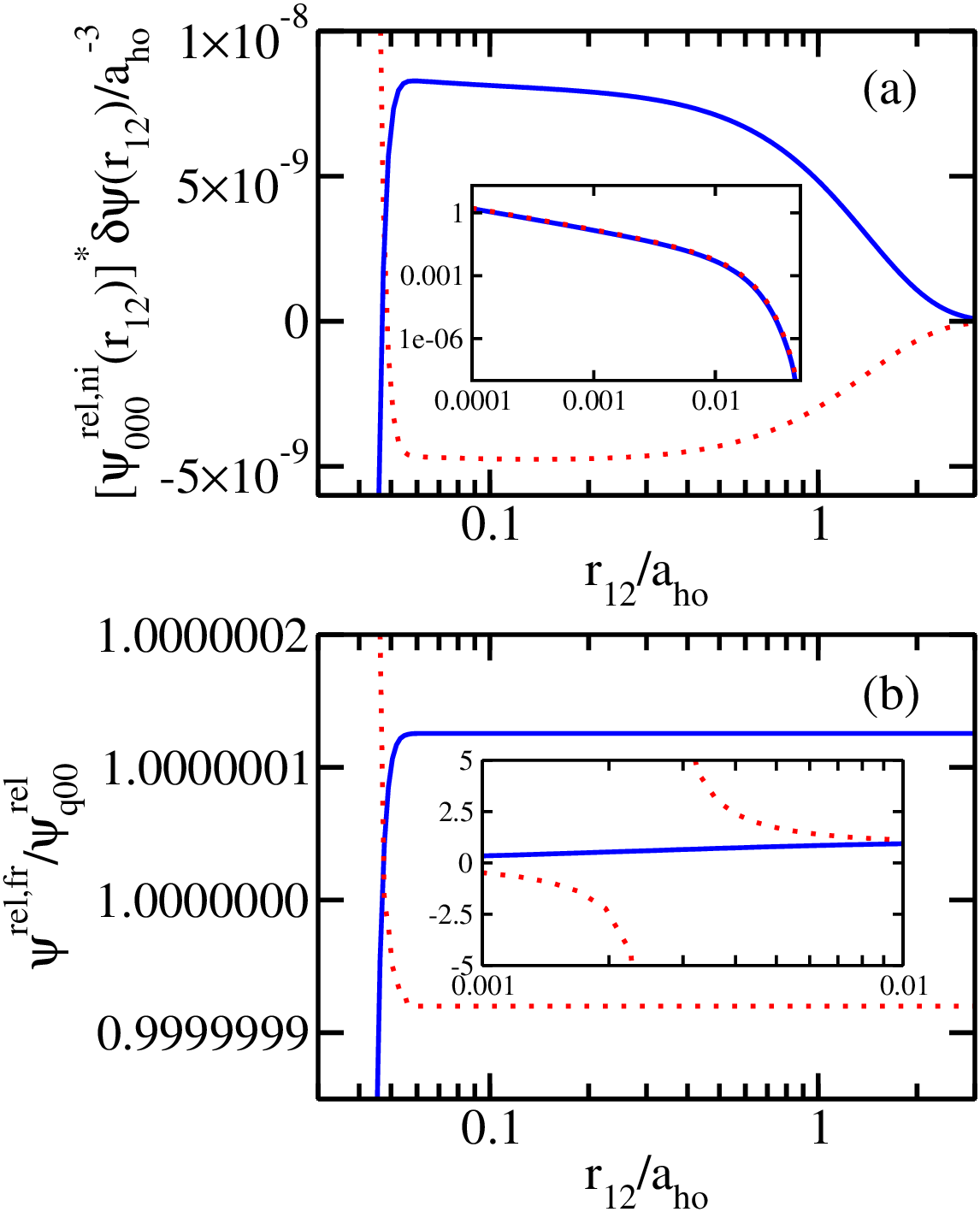}
\vspace*{0.2cm}
\caption{(Color online)
Analysis of the integrand that determines the non-universal
parameter $D_0$ [see
Eq.~(\ref{eq_u0trap})]
for the finite-range model potential 
$V_{\rm{g}}$ with $r_0=0.01a_{\rm{ho}}$
for two different relative two-body energies.
The solid and dotted lines correspond 
to $E_2^{\rm{rel}}=1.498 \hbar \omega$ 
and $E_2^{\rm{rel}}=1.502 \hbar \omega$, respectively.
Panel (a) shows the integrand
$[\psi_{000}^{\rm{rel,ni}}({\bf{r}}_{12})]^*\delta \psi({\bf{r}}_{12})$
in the large $r_{12}$ region, i.e.,
for $r_{12} \in [0.03a_{\rm{ho}},3a_{\rm{ho}}]$ (log-linear scale),
while the inset shows the absolute value of the integrand
$|[\psi_{000}^{\rm{rel,ni}}({\bf{r}}_{12})]^*\delta \psi({\bf{r}}_{12})|$
in the small 
$r_{12}$ region (log-log scale).
Panel (b) and the inset of (b) show the ratio
$\psi^{\rm{rel,fr}}/\psi_{q00}^{\rm{rel}}$ 
in the large and small $r_{12}$ regions, respectively (log-linear scale).
}\label{fig_wave}
\end{figure}

Figure~\ref{fig_wave} illustrates the behavior
of the integrand that determines the non-universal two-body
parameter $D_0$
for two different two-body energies,
i.e., for $E_2^{\rm{rel}}=1.498 \hbar \omega$ [negative $a_s(0)$]
and $E_2^{\rm{rel}}=1.502 \hbar \omega$ [positive $a_s(0)$],
for the Gaussian model potential with $r_0=0.01a_{\rm{ho}}$.
In particular, Fig.~\ref{fig_wave}(a)
shows the product
$(\psi_{000}^{\rm{rel,ni}})^*\delta \psi$
and 
Fig.~\ref{fig_wave}(b)
the ratio
$\psi^{\rm{rel,fr}} / \psi^{\rm{rel}}_{q00}$.
Figure~\ref{fig_wave} reflects the presence of the two 
characteristic length scales of the problem.
The behavior in the small $r_{12}$ region,
shown in the insets of Figs.~\ref{fig_wave}(a)
and \ref{fig_wave}(b), is governed by
the details of the two-body interaction potential.
Near $r_{12} \approx 5 r_0$, the behavior of 
$(\psi_{000}^{\rm{rel,ni}})^*\delta \psi$
and 
$\psi^{\rm{rel,fr}} / \psi^{\rm{rel}}_{q00}$
changes notably. 
For $r_{12} \gtrsim 5 r_0$,
the ratio 
$\psi^{\rm{rel,fr}}/\psi_{q00}^{\rm{rel}}$ 
approaches a constant that is slightly larger (smaller)
than 1 
for negative (positive) $a_s(0)$.
The small deviations of the ratios from one 
are a consequence of the fact that
the wave functions of the trapped system are normalized
to one. Since the wave functions for the finite
range interaction potential deviate from those for the
zero-range interaction potential in the
small $r_{12}$ region, the
ratio 
$\psi^{\rm{rel,fr}}/\psi_{q00}^{\rm{rel}}$ 
needs---in general---to
differ from one in the large $r_{12}$ region.
The ``divergence'' of the ratio 
$\psi^{\rm{rel,fr}}/\psi_{q00}^{\rm{rel}}$ 
near $r_{12}=0.002a_{\rm{ho}}$ for $a_s(0)>0$ 
[see dotted line in the inset of Fig.~\ref{fig_wave}(b)]
reflects the fact that the zero-range potential
supports a deeply-bound negative energy state, which introduces
a node at small $r_{12}$ in the wave function
that describes the energetically lowest-lying gas-like state.
A corresponding
bound-state is not supported by
the purely repulsive Gaussian model potential, leading to
an infinite ratio $\psi^{\rm{rel,fr}}/\psi_{q00}^{\rm{rel}}$ 
at the node of $\psi_{q00}^{\rm{rel}}$.

The non-universal two-body parameter $D_0$ is the trap analog of 
the two-body scattering quantity $u_0$ introduced
by Tan [see Eq.~(114a) of Ref.~\cite{tan08}].
It is important to note, however, that $D_0$ 
depends on the wave function difference for all $r_{12}$ 
and on the non-interacting harmonic oscillator
ground state wave function while 
the $u_0$ defined by Tan depends only on the wave function difference
in the small $r_{12}$ region, i.e., out to a few times $r_0$.
Indeed, we find that the contribution to
$D_0$ that accumulates in the inner region
($r_{12} \lesssim 10 r_0$) can be of comparable magnitude
to the contribution  that accumulates in the outer region
($r_{12} \gtrsim 10 r_0$).

\begin{figure}
\vspace*{+.9cm}
\includegraphics[angle=0,width=80mm]{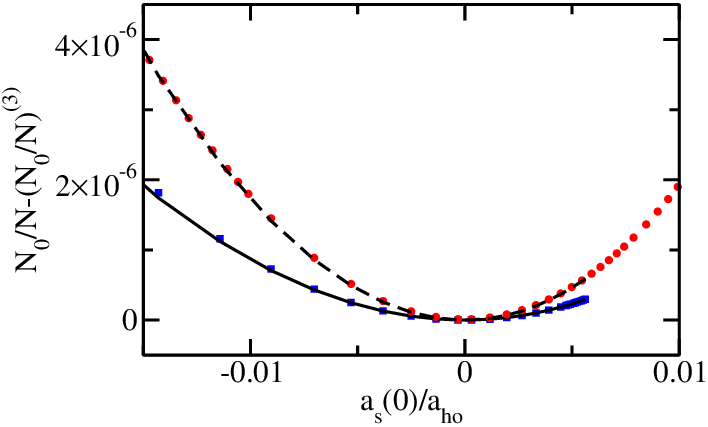}
\vspace*{0.2cm}
\caption{(Color online)
Squares and circles
show the residual
$N_0/N-(N/N_0)^{(3)}$
for $N=2$ and $3$
as a function of $a_s(0)$ for the Gaussian model
interaction $V_{\rm{g}}$ with $r_0=0.01 a_{\rm{ho}}$.
Solid and dashed lines show the quantity
$(N-1)  \delta c_{00}$
%^0$ 
for $N=2$ and $3$,
respectively~\cite{notation2}.
}\label{fig_condfracfinal}
\end{figure}

Since the dependence of the condensate fraction 
on the quantity $D_0$ 
arises
at the two-body level, the $D_0$ term needs
to be multiplied by $N-1$ for systems with $N>2$.
Figure~\ref{fig_condfracfinal} compares the condensate fractions for 
the $N=2$ and $3$ systems interacting through the 
finite-range Gaussian model potential $V_{\rm{g}}$
with the predicted behavior for the
condensate fraction.
Specifically,
squares and circles show the difference
$N_0/N - (N_0/N)^{(3)}$ for $N=2$ and $3$ between the 
numerically determined condensate fraction $N/N_0$
and
the perturbative result $(N_0/N)^{(3)}$, which
includes all terms on the right hand side
of Eq.~(\ref{eq_condpt}) up to order $a_{\rm{ho}}^{-3}$.
According to our discussion above, we expect that the
residuals are well approximated by 
$(N-1)  \delta c_{00}$
%^0$ 
(shown by solid and dashed lines in
Fig.~\ref{fig_condfracfinal}).
Indeed, Fig.~\ref{fig_condfracfinal} shows that
the residuals for $N=2$ and $3$ are well described by 
the non-universal corrections.
We find similar results for $N=4$ (not shown).
Our calculations demonstrate that
two Hamiltonians that are characterized by the same energy
give rise to condensate fractions that differ.
Related findings have previously been discussed in
Refs.~\cite{coes70,furn02,tan08}.
The leading order
difference between the condensate fractions 
for the harmonically trapped few-boson systems 
described by the two Hamiltonians
can be parameterized by the 
non-universal two-body parameter
$D_0$, a parameter not needed 
to match the energies of the two Hamiltonians.

\section{Three trapped bosons at unitarity}
\label{sec_threebody}
The previous section discussed the condensate fraction of 
weakly-interacting trapped $N$-boson systems, 
which can be expressed in terms
of $a_s(0)$, $r_e$ and a non-universal two-body correction
parametrized through $D_0$.
It is well known that the properties of the three-boson system
not only depend on two-body parameters, 
but
also on a three-body parameter~\cite{efim71,efim73,braa06}. 
In the weakly-interacting regime, however,
the dependence on the three-body parameter appears at
higher order 
than considered in Sec.~\ref{sec_nboson}.
In the strongly-interacting regime, in contrast, the dependence on the 
three-body parameter is generally quite pronounced.

At unitarity, 
i.e., for diverging $a_s(0)$,
the trapped three-boson system with zero-range $s$-wave
interactions supports two distinct classes of eigen states:
(i) universal states whose properties are fully governed by the
two-body scattering parameters, and 
(ii) non-universal states whose properties
depend, in addition to the two-body scattering
parameters, on a three-body parameter.
In the following, we determine the occupation numbers
for the non-universal three-boson states in
a trap at unitarity as a function
of the three-body parameter.
The momentum distribution of Efimov trimers
in free-space
was discussed in Ref.~\cite{wern11}.

The three-boson wave function $\psi({\bf{r}}_1,{\bf{r}}_2,{\bf{r}}_3)$
with relative orbital angular momentum $l=0$ for zero-range
interactions with diverging $s$-wave scattering length $a_s(0)$,
and vanishing $r_e$ and $V$,
under external isotropic harmonic confinement can be written 
as~\cite{jons02,wern06a}
\begin{eqnarray}
\label{eq_threebodywavefct}
\psi({\bf{r}}_1,{\bf{r}}_2,{\bf{r}}_3) = {\cal{S}}
\left[ R^{-5/2} F(R) \varphi(\alpha) \psi_{QLM}^{\rm{cm}}({\bf{R}}_{123})
\right].
\end{eqnarray}
Here, $R$ denotes the hyperradius and 
$\alpha$ the hyperangle,
$R^2 = r_{12}^2/2+2 r_{12,3}^2/3$ and 
$\tan \alpha = \sqrt{3}r_{12}/(2 r_{12,3})$
with $r_{12}=|{\bf{r}}_1-{\bf{r}}_2|$ and 
$r_{12,3}= | ({\bf{r}}_1+{\bf{r}}_2)/2-{\bf{r}}_3 |$.
In Eq.~(\ref{eq_threebodywavefct}), $\psi_{QLM}^{\rm{cm}}({\bf{R}}_{123})$
denotes the harmonic oscillator wave function
in the center of mass coordinate ${\bf{R}}_{123}$,
${\bf{R}}_{123}=({\bf{r}}_1+{\bf{r}}_2+{\bf{r}}_3)/3$;
as in Sec.~\ref{sec_twobody}, we assume that the center of mass
wave function is in the ground state, 
i.e., we set $Q=L=M=0$.
The operator ${\cal{S}}$ ensures that the three-boson wave function
is symmetric under the exchange of any of the three boson pairs,
${\cal{S}}=1+P_{12}+P_{23}+P_{31}+P_{12}P_{23}+P_{12}P_{31}$, 
where $P_{jk}$ is the operator
that exchanges particles $j$ and $k$.
The hyperangular wave function $\varphi(\alpha)$ takes the form
$\varphi(\alpha)=\sin[(\alpha-\pi/2)s_0]/\sin(2 \alpha)$,
where $s_0$ equals $1.00624  \imath$~\cite{efim71,efim73,braa06}. 
The fact that the separation 
constant $s_0$, which arises when solving the
hyperangular Schr\"odinger equation, is imaginary is unique
to the $l=0$ channel and 
tightly linked to the fact that 
a three-body
parameter is needed.

The hyperradial wave function $F(R)$ can be conveniently
expressed in terms of the Whittaker function $W$~\cite{wern06a},
i.e.,
$F(R)=R^{-1/2} W_{E^{\rm{rel}}_3/2,s_0/2}(R^2/a_{\rm{ho}}^2)$.
The relative three-body energy $E^{\rm{rel}}_3$ is related to the 
three-body or Efimov phase $\theta$ through~\cite{jons02}
\begin{eqnarray}
\label{eq_argthreebody}
\theta = \mbox{arg} \left(
\frac{\Gamma \left(\frac{1}{2}-\frac{E^{\rm{rel}}_3}{2\hbar \omega} + \frac{s_0}{2}
\right)}
{\Gamma(1+s_0)} \right).
\end{eqnarray}
The physical meaning of $\theta$ becomes clear
when looking at the small $R/a_{\rm{ho}}$
behavior of $F(R)$, 
$F(R) \rightarrow \sqrt{R} \sin ( \mbox{Im}(s_0) \ln(R/a_{\rm{ho}}) + \theta)$.
This expression shows that the three-body phase determines
what happens 
when three
particles come close together.
The small $R/a_{\rm{ho}}$ behavior can be thought of as 
being imposed by a short-range three-body force or a 
boundary condition of the hyperradial wave function
in the $R/a_{\rm{ho}} \rightarrow 0$ limit~\cite{braa06}.

To determine the occupation numbers of the non-universal three-boson
states as a function of $E^{\rm{rel}}_3$,
we sample the density $|\psi({\bf{r}}_1,{\bf{r}}_2,{\bf{r}}_3)|^2$
using Metropolis sampling. 
As discussed in Ref.~\cite{blum11}, this approach introduces
a statistical error that can be reduced by performing 
longer random walks.
Throughout our random walk, we
sample the projected one-body density matrix $\rho_{00}(r'_1,r_1)$.
Diagonalizing $\rho_{00}(r'_1,r_1)$ at the end of a run
yields the occupation numbers.

\begin{figure}
\vspace*{+.9cm}
\includegraphics[angle=0,width=60mm]{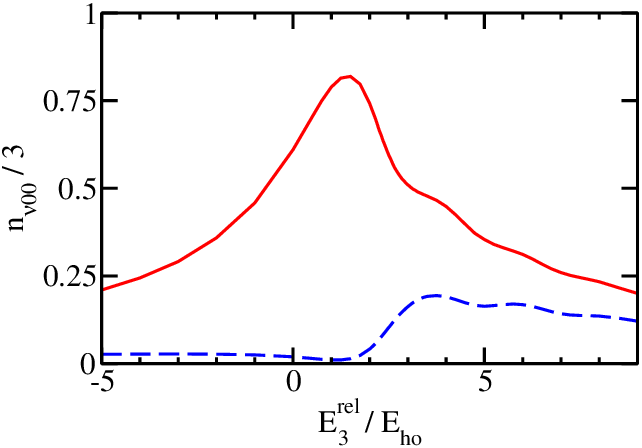}
\vspace*{0.2cm}
\caption{(Color online)
Occupation numbers per particle $n_{\nu00}/N$ 
[$\nu=0$ 
(solid line) and $\nu=1$ (dashed line)]
for the non-universal $l=0$ state
of the three-boson system
at unitarity as a function of the relative three-body
energy $E^{\rm{rel}}_3$ in units of the oscillator energy $E_{\rm ho}$,
$E_{\rm ho}=\hbar\omega$; the three-body energy can be converted to 
the three-body parameter via Eq.~(\ref{eq_argthreebody}).
}\label{fig_cond}
\end{figure}

Figure~\ref{fig_cond} shows the 
two largest occupation numbers per particle 
$n_{\nu00}/3$
as a function of $E_3^{\rm{rel}}$. 
It can be seen that the occupation numbers 
of the non-universal state depend quite strongly on the
relative three-body energy or, equivalently, the three-body phase $\theta$.
The maximum of the lowest occupation number per particle $n_{000}/3$ 
is $0.82$
and occurs at
$E_3^{\rm{rel}}=3 \hbar \omega/2$;
the occupation number per particle $n_{100}/3$  
is minimal at this energy.
Interestingly, the occupation numbers show oscillations 
(or ``shoulders'') similar to
those discussed in the context of Fig.~\ref{fig_occupationn2}
for the two-body system. 
In Figure~\ref{fig_occupationn2}, we change the 
relative two-body
energy, which is related to the $s$-wave scattering length 
through Eq.~(\ref{eq_energyn2}).
In Fig.~\ref{fig_cond}, we change the relative
three-body energy, which is related to the three-body phase through
Eq.~(\ref{eq_argthreebody}).
For both the two- and three-body systems, Eqs.~(\ref{eq_energyn2})
and (\ref{eq_argthreebody}) can be related to the short-range boundary
condition of the respective radial or hyperradial part of the 
relative wave function.

For comparison, we also calculated the largest occupation
number per particle of the projected one-body density matrix
$\rho_{00}(r'_1,r_1)$ for selected universal three-boson states
(see Ref.~\cite{wern06a} for the relevant wave functions).
The largest occupation number per particle of the energetically
lowest-lying universal
three-boson state with $l=0$ and
$E_3^{\rm{rel}}=4.465 \hbar \omega$ 
at unitarity is $n_{000}/3=0.295$.
The largest occupation numbers per particle
of the 
energetically lowest lying states with $l=1$ and $l=2$ 
at unitarity are $n_{000}/3=0.199$ and $0.421$, respectively.
For these states, the energies are $E_3^{\rm{rel}}=2.864 \hbar \omega$
and $E_3^{\rm{rel}}=2.823 \hbar \omega$, respectively.
For the universal states considered,
the largest occupation number $n_{000}$ 
is
notably smaller than $N$.

\section{Conclusions}
\label{sec_summary}
We have determined and interpreted the
occupation numbers of few-boson systems
under isotropic harmonic confinement.
In the weakly-interacting regime, our analysis is based on
a low-energy Hamiltonian---characterized by
the $s$-wave scattering length $a_s(0)$ and the effective
range $r_e$---that has previously been proven
to correctly describe the energy of few-boson
systems up to order $a_{\rm{ho}}^{-3}$~\cite{john12}.
The present paper shows that this low-energy
Hamiltonian correctly describes the leading order depletion
of harmonically trapped few-boson systems but that it
does not fully capture the corrections
to the leading order depletion.

Our final expression for the condensate fraction
reads
\begin{align}
\label{eq_condfracfinal}
N_0/N =1 & - 0.420004 (N-1)
\left[ \frac{a_s(0)}{a_{\rm{ho}}}\right]^2
\nonumber \\
& + \big[-0.373241
(N-1) \nonumber \\
& \quad\;\; + 0.439464
(N-1)(N-2)
\big] \left[\frac{a_s(0)}{a_{\rm{ho}}}\right]^3 \nonumber \\
& + \big[ 0.406786 (N-1) + \gamma_3^{(4)} (N-1)(N-2) \nonumber \\
& \quad\;\; + \gamma_4^{(4)} (N-1)(N-2)(N-3) \big]
\left[\frac{a_s(0)}{a_{\rm{ho}}}\right]^4 \nonumber \\
&+ 2 (N-1)\mbox{Re}(D_0) \nonumber \\
&- 
(3/2) \times 0.420004 (N-1)\frac{r_e [a_s(0)]^3}{a_{\rm{ho}}^4} \nonumber \\
&+ \cdots,
\end{align}
where the non-universal two-body parameter
$D_0$ 
is defined in Eq.~(\ref{eq_u0trap}).
The coefficients 
$\gamma_3^{(4)}$ and $\gamma_4^{(4)}$ 
arise when treating $V_{\rm{F}}$ in third-order
perturbation theory; the determination of their
numerical values is beyond the scope
of this paper.
We have confirmed 
the expression for the condensate fraction, Eq.~(\ref{eq_condfracfinal}),
through comparison with numerical results for
a few-body Hamiltonian with finite-range two-body potentials.
Our work demonstrates that the occupation numbers are not fully
determined by the parameters of the ``usual'' effective range
expansion, but rather depend 
on an additional property of the
two-body wave function (i.e., non-universal physics).
A similar result is expected to hold for the momentum distribution.
Our findings are not only of importance for cold atomic Bose
gases but also for nuclear systems, for which
the use of low-energy Hamiltonians has become increasingly more
popular during the past decade or so~\cite{epelbaum}.

We have also considered the strongly-interacting regime.
Our results show that the occupation numbers
for non-universal states
of the three-boson system under isotropic
harmonic confinement depend strongly on the three-body
parameter. This finding suggests that the occupation numbers
and momentum distribution
of strongly-interacting Bose gases at unitarity
may depend on three-body physics.
In view of recent experimental work~\cite{papp08,navo10,wild12},
it would be interesting to
extend the treatments of Refs.~\cite{song09,lee10},
which predict---accounting only for two-body physics---that 
three-dimensional Bose gases at unitarity fermionize.
In particular, it would be interesting to determine how, if at all,
this fermionization picture changes if three-body physics is 
accounted for.

\section{Acknowledgement}
Support by the NSF through
grant PHY-0855332,
and fruitful discussions with P. Johnson and E. Tiesinga
on the renormalized perturbation theory
framework are gratefully acknowledged.

\appendix

\section{Perturbative treatment and 
diagonalization of one-body density matrix}
\label{appendixA}
This appendix provides details regarding the perturbative treatment of 
the condensate fraction of the $N$-boson system and 
the diagonalization of the associated  matrix.

We start with the Hamiltonian $H$ given by Eq.~\eqref{eq_hamsecond}.
We first neglect the effective range dependent
potential $V'$, i.e., we 
consider only the bare Fermi pseudopotential $V_{\rm{F}}$,
Eq.~\eqref{eq_fermipseudo}, 
and the counterterm 
$W$
used to cure divergencies~\cite{john09,john12}.
The matrix elements 
$K_{\mathbf{abcd}}$ can then be written as
\begin{equation}
K_{\mathbf{abcd}}=
F_{\mathbf{abcd}} 
\left[
\frac{a_{s}(0)}{a_{\rm{ho}}}+\nu \left( \frac{a_{s}(0)}{a_{\rm{ho}}} \right)^2
\right] \hbar \omega ,
\end{equation}
where
\begin{equation}
F_{\mathbf{a}\mathbf{b}\mathbf{c}\mathbf{d}}=
4\pi 
a_{\rm{ho}}^3
\int 
\Phi_{\mathbf{a}}^{*}(\mathbf{r}_{1})
\Phi_{\mathbf{b}}^{*}(\mathbf{r}_{1}) 
\Phi_{\mathbf{c}}(\mathbf{r}_{1}) 
\Phi_{\mathbf{d}}(\mathbf{r}_{1}) d^{3} \mathbf{r}_{1}
\end{equation}
and
\begin{equation}
\nu=\sqrt{\frac{2}{\pi}}(1-\ln 2)+\sqrt{\frac{\pi}{2}} \beta_{2}^{(2)}.
\end{equation}
The coefficient $\beta_{2}^{(2)}$ was calculated in Ref.~\cite{john12} 
and is listed in Table~\ref{tab_coefficient}. It diverges
and cures the divergencies that arise when treating $V_{\rm{F}}$
in second-order perturbation theory.

The matrix 
$\langle \hat{a}_{\mathbf{p}}^{\dagger} \hat{a}_{\mathbf{q}} \rangle$ 
is evaluated 
by substituting Eq.~\eqref{eq_ptwavefunction} 
into Eq.~\eqref{eq_ptwavefunction_matrix}. 
In order to get the matrix elements up to order $a_{\rm{ho}}^{-3}$, we need
to employ second order perturbation theory. 
The 
expansion coefficients $b_{{\bf j}}^{(2)}$ 
of the non-normalized second-order
wave function $\psi_{\bf{0}}^{(2)}$ read
\begin{equation}
\label{secondorderptaaa}
b_{{\bf{0}}}^{(2)}=1
\end{equation}
and
\begin{align} 
\label{second order pt}
b_{{\bf j}}^{(2)} = & 
-\frac{\langle \psi_{{\bf j}}^{(0)}|V_{\rm{F}}+W|
\psi_{\bf{0}}^{(0)} \rangle}{E_{{\bf j}}^{(0)}-E_{\bf{0}}^{(0)}} 
\nonumber \\
& +\sum_{{\bf j}' \ne {\bf{0}}}
\frac{\langle \psi_{{\bf j}}^{(0)}|V_{\rm{F}}+W|
\psi_{{\bf j}'}^{(0)} \rangle 
\langle \psi_{{\bf j}'}^{(0)}|V_{\rm{F}}+W|\psi_{\bf{0}}^{(0)} \rangle}
{(E_{{\bf j}}^{(0)}-E_{\bf{0}}^{(0)})(E_{{\bf j}'}^{(0)}-E_{\bf{0}}^{(0)})} 
\nonumber \\
& - \frac{\langle \psi_{{\bf j}}^{(0)}|V_{\rm{F}}+W|
\psi_{\bf{0}}^{(0)} \rangle 
\langle \psi_{\bf{0}}^{(0)}|V_{\rm{F}}+W|\psi_{\bf{0}}^{(0)} \rangle}
{(E_{{\bf j}}^{(0)}-E_{\bf{0}}^{(0)})^{2}}
\end{align}
for ${\bf{j}}$ not equal to the ground state
labeled by
${\bf{0}}$.
In the denominators appearing in Eq.~(\ref{second order pt}),
the $E_{\bf{j}}^{(0)}$ denote the unperturbed eigen energies
corresponding to the ${\bf{j}}$'s unperturbed eigen state.
The numerators are conveniently expressed 
in terms of
the matrix elements 
$F_{\mathbf{a}\mathbf{b}\mathbf{c}\mathbf{d}}$.

The indices $\mathbf{p}$ and 
$\mathbf{q}$ of $\langle \hat{a}_{\mathbf{p}}^{\dagger} 
\hat{a}_{\mathbf{q}} \rangle$ 
run over all possible single particle state labels. 
We employ spherical coordinates and
write ${\bf{p}}=(n_1',l_1',m_1')$ and
${\bf{q}}=(n_1,l_1,m_1)$.
We find that the matrix is block
diagonal, i.e.,
 $\langle \hat{a}_{n_1' l_1' m_1'}^{\dagger} \hat{a}_{n_1 l_1 m_1} \rangle=0$
for $l_1' \ne l_1$ or $m_1' \ne m_1$.
In the following, we consider the submatrix
with $l_1'=l_1=m_1'=m_1=0$. 
We denote the matrix elements by $c_{n_{1}'n_{1}}$
and write
\begin{equation}
c_{n_{1}'n_{1}}
=\sum_{k=0}^{3}c_{n_{1}'n_{1}}^{(k)}
x^{k} + {\cal{O}}(x^4),
\end{equation}
where $x = a_s(0)/a_{\rm{ho}}$.
Considering symmetry
and keeping terms up to order $x^3$, 
we find
\begin{widetext}
\begin{equation}
\begin{split}
\langle 
  \hat{a}_{n_{1}'00}^{\dagger} \hat{a}_{n_{1}00} \rangle= 
\left( \begin{array}{cccc}
1+c_{00}^{(2)}x^2+c_{00}^{(3)}x^3 & c_{10}^{(1)}x+c_{10}^{(2)}x^2+c_{10}^{(3)}x^3 & \cdots & c_{A0}^{(1)}x+c_{A0}^{(2)}x^2+c_{A0}^{(3)}x^3\\
c_{10}^{(1)}x+c_{10}^{(2)}x^2+c_{10}^{(3)}x^3 & c_{11}^{(2)}x^2+c_{11}^{(3)}x^3 & \cdots & c_{A1}^{(2)}x^2+c_{A1}^{(3)}x^3\\
\vdots & \vdots & \ddots & \vdots\\
c_{A0}^{(1)}x+c_{A0}^{(2)}x^2+c_{A0}^{(3)}x^3 & c_{A1}^{(2)}x^2+c_{A1}^{(3)}x^3 & \cdots & c_{AA}^{(2)}x^2+c_{AA}^{(3)}x^3 \end{array} \right).
\end{split}
\end{equation}
\end{widetext}
The upper left element is $1$, with small corrections 
proportional to
$x^2$ and $x^3$.
The leading-order contribution of the other elements 
in the first row and first column is proportional to
$x$. The leading-order contribution of 
the rest of the matrix elements is proportional to
$x^2$. 

We diagonalize the matrix by solving
\begin{equation}
\label{eq_determinant}
{\mbox{det}}( \mathbf{M})= {\mbox{det}} (\langle \hat{a}_{n_{1}' 00}^{\dagger} 
\hat{a}_{n_{1} 00} \rangle-\Xi \mathbf{I} )=0
\end{equation}
through application of 
the Leibniz formula for determinants~\cite{weber}. 
In Eq.~(\ref{eq_determinant}),
${\bf{I}}$ denotes the $(A+1) \times (A+1)$ 
identity matrix
and $\Xi$ the eigen value we are seeking.
The product of the diagonal elements
%, i.e.,
%the term without permutations, 
can be written as
\begin{align}
\label{eq_diagonal}
\prod_{i=1}^{A+1}M_{ii} = 
& (-\Xi)^{A+1}+\Bigg[ 1+
\sum_{j=0}^{A}
\left( c_{jj}^{(2)}x^2+c_{jj}^{(3)}x^3 \right) 
\Bigg](-\Xi)^{A} \nonumber \\
& + \sum_{j=1}^{A}\left( c_{jj}^{(2)}x^2+c_{jj}^{(3)}x^3 \right)(-\Xi)^{A-1} 
 +\mathcal{O}\left(x^4\right).
\end{align}
The other terms involve the product of the
diagonal elements
with the first and $k$$^{th}$ diagonal elements replaced
by $M_{1k}$ and $M_{k1}$.
For $k=2$, for example, we have
%If 
%$\sigma_{1k}$ 
%denotes the operation that exchanges the indices ``$1$'' and
%``$k$'',
%we find
\begin{align}
& \frac{M_{12}M_{21}}{M_{11} M_{22}}\prod_{i=1}^{A+1}M_{ii} = \nonumber \\
& \quad \left( c_{10}^{(1)}x+c_{10}^{(2)}x^2+c_{10}^{(3)}x^3 \right)^{2} 
%\nonumber \times \\
\prod_{j=2}^A \left( c_{jj}^{(2)}x^2+c_{jj}^{(3)}x^3 -\Xi \right) 
\nonumber \\
& = \left[ (c_{10}^{(1)})^{2}x^{2}+2c_{10}^{(1)}c_{10}^{(2)}x^{3} \right] (-\Xi)^{A-1}
 +\mathcal{O}\left(x^4\right). 
\end{align}
Summing over all contributions with $k \ge 2$, we find
\begin{align}
\label{eq_permute2}
\sum_{k=2}^{A+1} & \frac{M_{1k}M_{k1}}{M_{11} M_{kk}}\prod_{i=1}^{A+1}M_{ii} = \nonumber \\
& \sum_{j=1}^{A}\left[ (c_{j0}^{(1)})^{2}x^{2}+2c_{j0}^{(1)}c_{j0}^{(2)}x^{3} \right] (-\Xi)^{A-1}
+\mathcal{O}\left(x^4\right).
\end{align}
Combining Eqs.~(\ref{eq_diagonal}) and (\ref{eq_permute2})
yields the eigen value equation up to order $x^3$,
\begin{align}
\label{eq_eigenfinal}
(-\Xi)^{A+1}&+\Bigg[ 1+
\sum_{j=0}^{A}
\left( 
c_{jj}^{(2)}x^2+c_{jj}^{(3)}x^3 \right) \Bigg](-\Xi)^{A} \nonumber \\
& + \sum_{j=1}^{A}\bigg[ c_{jj}^{(2)}x^2+c_{jj}^{(3)}x^3+(c_{j0}^{(1)})^{2}x^{2} \nonumber \\
& \qquad \; \; \; + 2c_{j0}^{(1)}c_{j0}^{(2)}x^{3} \bigg](-\Xi)^{A-1}=0.
\end{align}
Equation~(\ref{eq_eigenfinal}) can be reduced to a quadratic equation
in $\Xi$. 
Taking $A$ to infinity,
the largest eigen value coincides with 
the condensate fraction,
\begin{align} 
\label{eq_cf_order2and3}
N_{0}/N = 1 &+ \left[c_{00}^{(2)}+\sum_{j=1}^{\infty}(c_{j0}^{(1)})^{2}\right]x^2 \\
& +\left[c_{00}^{(3)}
+\sum_{j=1}^{\infty}2c_{j0}^{(1)}c_{j0}^{(2)}\right]x^3+\mathcal{O}\left(x^4\right). \nonumber 
\end{align}
The coefficients 
$c_{mn}^{(k)}$ are determined by 
Eqs.~\eqref{eq_ptwavefunction_matrix},
\eqref{eq_ptwavefunction},  
\eqref{secondorderptaaa}, 
and  \eqref{second order pt},
and can be expressed in terms of infinite sums involving the matrix elements 
$F_{\mathbf{a}\mathbf{b}\mathbf{c}\mathbf{d}}$
(see Table~\ref{tab_coefficient}). 
Evaluating the coefficients $c_{mn}^{(k)}$,
Eq.~\eqref{eq_cf_order2and3}
becomes
\begin{align}
\label{eq_condfracgamma}
N_{0}/N = 1 & - \gamma_{2}^{(2)}(N-1)
\left[\frac{a_{s}(0)}{a_{\rm{ho}}}\right]^{2} \nonumber \\
& + \bigg[\gamma_2^{(3)} (N-1) \nonumber \\
& \quad \;\; + \gamma_3^{(3)} (N-1)(N-2) \bigg]
\left[\frac{a_{s}(0)}{a_{\rm{ho}}}\right]^{3},
\end{align}
where 
$\gamma_2^{(3)}= -2\gamma_{2,1}^{(3)}-4\gamma_{2,2}^{(3)}-2\gamma_{2,3}^{(3)}$
and
$\gamma_3^{(3)}= 
-4\gamma_{3,1}^{(3)}-4\gamma_{3,2}^{(3)}+8\gamma_{3,3}^{(3)}+4\gamma_{3,4}^{(3)}$.
The superscript and the first subscript
of the coefficient $\gamma_{i,j}^{(k)}$ denote
respectively the orders of $a_s(0)/a_{\rm{ho}}$ 
and the 
multi-body scattering process
that $\gamma_{i,j}^{(k)}$ 
is associated with.
The second subscript simply labels the various sums (see
Table~\ref{tab_coefficient}).
To evaluate $\gamma_{3,4}^{(3)}$,
we use the expression
\begin{align}
\label{eq_gamma343}
\gamma_{3,4}^{(3)} = &
\left(\frac{2}{\pi}\right)^{3/2} \left[
\frac{\pi^2}{24} + \ln 2 - \frac{1}{2} \ln^2 2 \right] \times \nonumber \\
& \quad \left[\sqrt{\frac{4}{3}} + \ln ( 8 - 4 \sqrt{3}) -1
\right] \nonumber \\
& -  
\sum_{j=1}^{\infty}\sum_{k=1}^{\infty} 
\frac{2^{1/2-2j-2k} \Gamma(j+k+3/2)}{j^2 k \pi^2 j! k! }.
\end{align}
If we insert the numerical values of the coefficients 
$\gamma_{i,j}^{(k)}$
from Table~\ref{tab_coefficient}, we obtain Eq.~(\ref{eq_condpt})
of the main text.

To understand how the effective range contributes
to the depletion of the condensate fraction, we treat the 
potential $V'$ in first-order perturbation theory. 
We find that the $V'$ does not give rise to a term proportional to
$r_e [a_s(0)]^2 / a_{\rm{HO}}^3$.

{
\begin{table}
\caption{Expressions for and numerical values
of the coefficients 
$\gamma_{i,j}^{(k)}$ that enter into 
Eq.~(\protect\ref{eq_condfracgamma}).
The representation of the $\gamma_{i,j}^{(k)}$
in terms of infinite sums, derived within the
perturbative framework, are listed in column 2.
For completeness, we also list the coefficient
$\beta_2^{(2)}$, which enters into the counterterm $W$
needed to cure the divergencies arising from $V_{\rm{F}}$.
$\Delta \epsilon_{{\bf{a}}{\bf{b}}}$ denotes
a dimensionless energy; in spherical coordinates,
we have
$\Delta \epsilon_{{\bf{a}}{\bf{b}}}=2n_a+l_a+2n_b+l_b$.
The sums are over all vector indices 
with the restrictions 
$\mathbf{a} \neq \mathbf{0}$, $\mathbf{b} \neq \mathbf{0}$ 
and $\mathbf{c} \neq \mathbf{0}$
(e.g., the sum that determines
$\gamma_2^{(2)}$ is
$\sum = \sum_{{\bf{a}} \ne {\bf{0}},{\bf{b}} \ne {\bf{0}}}$,
where ${\bf{a}}={\bf{0}}$ corresponds to $n_a=l_a=0$).
The numerical values for the coefficients are given in column 3:
$\gamma_{2}^{(2)}$, $\gamma_{2,1}^{(3)}$, $\gamma_{2,2}^{(3)}$,
$\gamma_{2,3}^{(3)}$ and $\gamma_{3,4}^{(3)}$
are obtained by evaluating 
Eqs.~(\protect\ref{eq_gamma22}), 
(\protect\ref{eq_gamma213}), (\protect\ref{eq_gamma223}),
(\protect\ref{eq_gamma233}), and (\protect\ref{eq_gamma343})
while
$\gamma_{3,3}^{(3)}$ is
obtained by evaluating the infinite sum numerically
(the numerical uncertainty is reported in round brackets).
In terms of the $\alpha$ coefficients defined in Ref.~\cite{john12},
we have 
$\gamma_2^{(2)}=(\alpha_{4,3}^{(3)}-2 \alpha_5^{(3)})/\alpha_2^{(1)}$.
}
\begin{ruledtabular}
\begin{tabular}{c|c|l}
   & infinite sum &  \multicolumn{1}{c}{numerical value}\\
  \hline
  $\beta_{2}^{(2)}$ 
  & $\sum\frac{F_{\mathbf{00ab}}F_{\mathbf{ba00}}}{\Delta \epsilon_{\mathbf{ab}}}+2 
  \sum\frac{F_{\mathbf{000a}}F_{\mathbf{a000}}}{\Delta \epsilon_{\mathbf{a0}}} $  
  & \multicolumn{1}{c}{diverges}
   \\
  \hline
  $\gamma_{2}^{(2)}$ 
  & $\sum\frac{F_{\mathbf{00ab}}F_{\mathbf{ba00}}}{\Delta \epsilon_{\mathbf{ab}}^2}$  
  & 0.420004291120 \\
  \hline
  $\gamma_{2,1}^{(3)}$ 
  & $F_{\mathbf{0000}}\sum\frac{F_{\mathbf{00ab}}F_{\mathbf{ba00}}}
  {\Delta \epsilon_{\mathbf{ab}}^3}$
  & 0.073250101788 \\
  \hline
  $\gamma_{2,2}^{(3)}$ 
  & $\sum\frac{F_{\mathbf{000a}}F_{\mathbf{a00b}}F_{\mathbf{b000}}}
  {\Delta \epsilon_{\mathbf{a0}}\Delta \epsilon_{\mathbf{ab}}^2}$ 
  & 0.005269765990 \\
  \hline
  $\gamma_{2,3}^{(3)}$ 
  & $(1-\ln 2)\sqrt{\frac{2}{\pi}} \gamma_2^{(2)}$
  & 0.102830963978 \\
  \hline
  $\gamma_{3,1}^{(3)}$ 
  & $\gamma_{2,1}^{(3)}$ 
  & 0.073250101788 \\
  \hline
  $\gamma_{3,2}^{(3)}$  
  & $\gamma_{2,2}^{(3)}$ 
  & 0.005269765990 \\
  \hline
  $\gamma_{3,3}^{(3)}$ 
  & $\sum\frac{F_{\mathbf{00ab}}F_{\mathbf{b00c}}F_{\mathbf{ca00}}}
  {\Delta \epsilon_{\mathbf{ac}}\Delta \epsilon_{\mathbf{ab}}^2}$ 
  & 0.067074(1) \\
  \hline
  $\gamma_{3,4}^{(3)}$ 
  & $\sum\frac{F_{\mathbf{000a}}F_{\mathbf{a0bc}}F_{\mathbf{cb00}}}
  {\Delta \epsilon_{\mathbf{a0}}\Delta \epsilon_{\mathbf{bc}}^2}$ 
  & 0.054238116273 \\
\end{tabular}
\end{ruledtabular}
\label{tab_coefficient}
\end{table}

\section{Determination of one-body density matrix for $N=2$}
\label{appendixB}
This appendix summarizes the
evaluation of the one-body density matrix for the two-boson
system with regularized
$\delta$-function interaction
in a spherically symmetric harmonic trap. 
We start with Eq.~(\ref{eq_onebody}) 
and write the two-body wave function as a product
of the center-of-mass wave function
$\psi_{QLM}^{\rm{cm}}({\bf{R}}_{12})$ and the relative
wave function $\psi_{qlm}^{\rm{rel}}({\bf{r}}_{12})$. 
In the following, we assume that the 
two-body wave function is normalized and
restrict ourselves to states
with $Q=L=M=l=m=0$, yielding
\begin{align}
\label{eq_step1}
\rho({\bf r}_1',{\bf r}_1) = 2 \int & 
[\psi^{\rm{cm}}_{000}({\bf R}_{12}') 
\psi^{\rm{rel}}_{q00}({\bf r}_{12}')]^* 
\times \nonumber \\
& \psi^{\rm{cm}}_{000}({\bf R}_{12}) 
\psi^{\rm{rel}}_{q00}({\bf r}_{12})d^3{\bf r}_2.
\end{align}
To evaluate Eq.~(\ref{eq_step1}), we follow a three-step  process:
(i) We expand the relative wave function in terms of a complete set of
non-interacting harmonic oscillator wave functions in the relative
coordinates.
(ii) We expand the non-interacting relative and center of mass wave
functions
in terms of non-interacting single particle harmonic oscillator 
wave functions.
(iii) We integrate over ${\bf{r}}_2$.

Step (i):
The relative wave function
reads~\cite{busc98}
\begin{align}
\label{eq_psi_r12INT}
\psi^{\rm{rel}}_{q00}({\bf{r}}_{12}) & = 
\frac{N_q^{\rm{rel}}}{\sqrt{4 \pi}} 
U\!\!\left(\!-q,\tfrac{3}{2},\tfrac{1}{2}\!\! 
\left[ \tfrac{r_{12}}{a_{\rm{ho}}}\right]^2 \right) 
e^{-\tfrac{1}{4} \left(\tfrac{r_{12}}{a_{\rm{ho}}}\right)^2},
\end{align}
where $U$ is the confluent hypergeometric function
and the normalization constant $N_q^{\rm{rel}}$ is given by
\begin{align}
\label{eq_psi_r12INTnorm}
N_q^{\rm{rel}} = 
\sqrt{\frac{2^{2q} \; \Gamma(-1-2q) \; a_{\rm{ho}}^{-3} \; \pi^{-1/2} \; 2^{3/2}}
{1/q + \pi \cot(\pi q) - \psi(-q-1/2) + \psi(q)}}.
\end{align}
Here, $\psi$ is the digamma function
and the non-integer quantum number $q$ is determined
by the $s$-wave scattering length via 
Eqs.~(\ref{eq_defineq}) and (\ref{eq_energyn2}).
In the non-interacting limit, we have
\begin{align}
\label{eq_psi_r12NI}
\psi^{\rm{rel,ni}}_{i00}({\bf r}_{12}) & = \frac{N_i^{\rm{rel,ni}}}{\sqrt{4 \pi}} 
L_i^{(1/2)}\!\!\left(\tfrac{1}{2}\!\left[\frac{r_{12}}{a_{\rm{ho}}}\right]^2\right)
e^{-\tfrac{1}{4} \left(\tfrac{r_{12}}{a_{\rm{ho}}}\right)^2}
\end{align}
with
\begin{align}
\label{eq_psi_r12NInorm}
N_i^{\rm{rel,ni}} = 
\sqrt{\frac{i! \; a_{\rm{ho}}^{-3}}{\Gamma(i+3/2)\sqrt{2}}}.
\end{align}
In Eq.~(\ref{eq_psi_r12NI}), 
the $L_i^{(1/2)}$ denote the associated Laguerre polynomials.
Using the generating function of the confluent hypergeometric 
function~\cite{busc98},
\begin{align}
\label{eq_Ugenfct}
\Gamma(-q) U\!\!\left(-q,\tfrac{3}{2},x\right) = \sum_{i=0}^{\infty}
\frac{L_i^{(1/2)}(x)}{i-q},
\end{align}
the interacting wave function $\psi^{\rm{rel}}_{q00}({\bf{r}}_{12})$
can be expanded in terms of the non-interacting 
wave functions $\psi^{\rm{rel,ni}}_{i00}({\bf{r}}_{12})$,
\begin{eqnarray}
\label{eq_wavefctrel}
\psi_{q00}^{\rm{rel}}({\bf{r}}_{12}) = \sum_{i=0}^{\infty} 
C_i
\psi^{\rm{rel,ni}}_{i00}({\bf{r}}_{12}) ,
\end{eqnarray}
where
\begin{eqnarray}
\label{eq_coeffpseudo}
C_i
=\frac{N_q^{\rm{rel}}}{N_i^{\rm{rel,ni}} \Gamma(-q) (i-q)}. 
\end{eqnarray}
Inserting the right hand side of 
Eq.~(\ref{eq_wavefctrel}) into Eq.~(\ref{eq_step1}),
the one-body density matrix reads
\begin{align}
\label{eq_step2}
\rho({\bf r}_1', & {\bf r}_1) = 2 \sum_{i=0}^{\infty} 
\sum_{i'=0}^{\infty} 
C_{i'}^{*} C_{i} \times \\
&\int [\psi^{\rm{cm}}_{000}({\bf R}_{12}')
\psi^{\rm{rel,ni}}_{i'00}({\bf r}_{12}')]^* 
\psi^{\rm{cm}}_{000}({\bf R}_{12}) \psi^{\rm{rel,ni}}_{i00}({\bf r}_{12})d^3{\bf r}_2. \nonumber
\end{align}

Step (ii):
To facilitate the integration over ${\bf{r}}_2$ in Eq.~(\ref{eq_step2}),
we expand the product of the non-interacting relative and center
of mass wave functions in terms of single particle states,
\begin{eqnarray}
\label{eq_transformation}
& \sum_{Mm} \psi_{QLM}^{\rm{cm}}({\bf{R}}_{12}) \psi_{ilm}^{\rm{rel,ni}}({\bf{r}}_{12}) \langle L,M,l,m|\Lambda,\Pi \rangle = 
\nonumber \\ 
& \sum_{n_1 l_1 m_1} \sum_{n_2 l_2 m_2} 
\langle \langle n_1, l_1, n_2, l_2 ; \Lambda| Q,L,i,l; \Lambda \rangle \rangle \times
\nonumber \\
& \langle l_1,m_1, l_2, m_2 | \Lambda,\Pi\rangle 
\Phi_{n_1 l_1 m_1}({\bf{r}}_1) 
\Phi_{n_2 l_2 m_2}({\bf{r}}_2),
\end{eqnarray}
where $\langle \langle \cdots \rangle \rangle$ denotes
a Talmi-Moshinsky coefficient~\cite{talmi,moshinsky},
$\langle \cdots \rangle$ a Clebsch Gordon coefficient
and $\Phi_{nlm}({\bf{r}})$ the single particle harmonic oscillator 
wave function,
\begin{align}
\label{eq_spWF}
\Phi_{nlm}({\bf{r}}) = 
R_{nl}(r) Y_{lm}(\hat{r})
\end{align}
with
\begin{align}
R_{nl}(r) = 
N_{nl}^{\rm{sp}} \! \left(\frac{r}{a_{\rm{ho}}}\right)^{\!\!l}
\! L_{n}^{(l+1/2)}\!\!\left(\frac{r^2}{a_{\rm{ho}}^2}\right)
\! e^{-\tfrac{1}{2}\left(\tfrac{r}{a_{\rm{ho}}}\right)^2} 
\end{align}
and
\begin{align}
N_{nl}^{\rm{sp}} = \sqrt{\frac{2 \; n! \; a_{\rm{ho}}^{-3}}
{\Gamma(n + l + 3/2)}}.
\end{align}
In Eq.~(\ref{eq_transformation}), $\Lambda$ 
denotes the total angular momentum quantum number to which the two-particle
state on the left hand side is coupled
and $\Pi$ the corresponding projection quantum number.
For the state of interest,
we have $\Lambda=0$ since $L=l=0$. Correspondingly, we
have $\Pi=0$. This implies that the sums on the left hand side of
Eq.~(\ref{eq_transformation}) reduce to a single term
with Clebsch-Gordon coefficient
$\langle 0,0,0,0|0,0 \rangle =1$.
For $\Lambda=\Pi=0$, the Clebsch-Gordon coefficient
on the right hand side of Eq.~(\ref{eq_transformation})
is only non-zero if $l_2=l_1$ and $m_2=-m_1$, which
eliminates the sums over $l_2$ and $m_2$ and yields
$\langle l_1,m_1,l_1,-m_1|0,0 \rangle = (-1)^{l_1 - m_1} (2l_1+1)^{-1/2}$.
Using these constraints for the quantum numbers,
the Talmi-Moshinsky bracket 
on the right hand side of Eq.~(\ref{eq_transformation})
reduces to~\cite{trlifaj}
\begin{align}
\label{eq_TMcoeff}
\langle \langle n_1, l_1, & n_2, l_1;0|0,0,i,0;0 \rangle \rangle = \nonumber \\
& \frac{(-1)^{l_1}}{2^i} \sqrt{2 l_1+1} \frac{i!}{n_1! \; n_2!} 
\frac{N_{n_1l_1}^{\rm{sp}}N_{n_2l_1}^{\rm{sp}}}
{N_{00}^{\rm{sp}} N_{i0}^{\rm{sp}} }. 
\end{align}
Energy conservation implies that $i$ is constrained to take the values
$i=n_1+n_2+l_1$ in Eq.~(\ref{eq_TMcoeff}).
Applying Eq.~(\ref{eq_transformation}) twice to the integrand of 
Eq.~(\ref{eq_step2}), with the associated restrictions
on the quantum numbers, we find
\begin{align}
\label{eq_step3}
\rho({\bf r}_1', & {\bf r}_1) = 
2 \sum_{n_1' l_1' m_1' n_2'} \sum_{n_1 l_1 m_1 n_2} 
[(2 l_1 +1)(2 l_1' +1)]^{-1/2}
\times \nonumber \\
& (C_{n_1'+n_2'+l_1'})^{*} C_{n_1+n_2+l_1} \; (-1)^{l_1-m_1+l_1'-m_1'}\times \nonumber \\
& \langle \langle n_1, l_1, n_2, l_1;0|0,0,n_1+n_2+l_1,0;0 \rangle \rangle \; \times \nonumber \\
& \langle \langle n_1', l_1', n_2', l_1';0|0,0,n_1'+n_2'+l_1',0;0 \rangle \rangle  \; \times \nonumber \\
& 
[\Phi_{n_1'l_1'm_1'}({\bf r}_1')]^* \Phi_{n_1l_1m_1}({\bf r}_1) 
\nonumber \times \\
& \int [\Phi_{n_2'l_1'-m_1'}({\bf r}_2)]^* 
\Phi_{n_2l_1-m_1}({\bf r}_2)
d^3{\bf r}_2,
\end{align}
where the sums over $i$ and $i'$ have been eliminated due to
the energy conservation constraint.

Step (iii):
The integration over ${\bf{r}}_2$ only gives non-vanishing contributions
if $n_2'=n_2$, $l_1'=l_1$ and $m_1'=m_1$.
We thus obtain
\begin{align}
\label{eq_OBDM}
\rho({\bf r}_1',{\bf r}_1) & = 
2 
\sum_{n_1' l_1 m_1} \sum_{n_1}
c_{n_1'n_1}^{l_1} 
[\Phi_{n_1'l_1m_1}({\bf r}_1')]^* 
\Phi_{n_1 l_1m_1}({\bf r}_1),
\end{align}
where
\begin{align}
\label{eq_cn1n1p}
c_{n_1'n_1}^{l_1} = & \sum_{n_2=0}^{\infty} 
\frac{(C_{n_1'+n_2+l_1})^{*} C_{n_1+n_2+l_1}}{2 l_1+1}\times \nonumber \\
& 
\langle \langle n_1, l_1, n_2, l_1;0|0,0,n_1+n_2+l_1,0;0 \rangle \rangle \times \nonumber \\
& \langle \langle n_1', l_1, n_2, l_1;0|0,0,n_1'+n_2+l_1,0;0 \rangle \rangle.
\end{align}

The projected one-body density matrix
$\rho_{\lambda \mu}(r_1',r_1)$,
Eq.~(\ref{eq_projection}), can now be calculated readily.
In the following we consider the case where $\lambda =0$
and
drop the superscript 
of $c_{n_1' n_1}^{l_1}$
for notational convenience.
We find
\begin{align}
\label{eq_OBDMprojected}
\rho_{00}(r_1',r_1) = &
\; 2 \sum_{n_1'n_1} c_{n_1'n_1} R_{n_1'0}(r_1') R_{n_10}(r_1),
\end{align}
where the $c_{n_1' n_1}$ 
can be interpreted as elements of a symmetric coefficient matrix
whose eigen values are the scaled occupation
numbers $n_{\nu 00}/2$.  The 
$n_{\nu 00}/2$ are
shown in Fig.~\ref{fig_occupationn2}.

In the weakly-interacting regime, we obtain analytic expressions
for the occupation numbers of the ground state
by expanding around $q=0$.
Using Eq.~\eqref{eq_energyn2} with $r_e=0$, 
we rewrite the $c_{n_1' n_1}$ in terms 
of 
$x=a_s(0)/a_{\rm{ho}}$ as opposed to $q$.
Our goal is to obtain the condensate fraction
of the weakly-interacting two-body ground state up to
fourth order in $x$.
Extending the analytical procedure discussed in Appendix~\ref{appendixA},
this requires that we calculate
$c_{00}$ up to fourth order in $x$, $c_{j0}$ up to third order, and
$c_{jj}$ up to second order. 
Inspection of Eq.~(\ref{eq_cn1n1p})
shows that the $a_s(0)$-dependence of $c_{n_1' n_1}$ comes from
the $C_{n_1+n_2}$ and $(C_{n_1'+n_2})^{*}$ coefficients.
We write
\begin{align}
\label{eq_Cq_i0_fourthOrder}
C&_j \approx C_j^{(0)} + C_j^{(1)}x +C_j^{(2)}x^2 + C_j^{(3)}x^3 + C_j^{(4)}x^4 + {\cal O}(x^5).
\end{align}
The $C_j^{(k)}$'s needed to evaluate the condensate fraction up to order $x^4$
are
\begin{align}
\label{eq_Cq_coeff1}
C_0^{(0)}&=1,\\
\label{eq_Cq_coeff2}
C_0^{(1)}&=0,\\
C_0^{(2)}&=\frac{1}{8 \pi} \left( h_{0,2}+h_{0,1}^2 \right), \\
C_0^{(3)}&=- \frac{1}{3(2 \pi)^{3/2}} \left(h_{0,3} + 3 h_{0,2}h_{0,1} +2h_{0,1}^3\right),
\end{align}
and
\begin{align}
C_0^{(4)}&=\frac{1}{128 \pi^2} \big(12 h_{0,4} +48 h_{0,3} h_{0,1} + 17 h_{0,2}^2
\nonumber \\
& \qquad + 106 h_{0,2}h_{0,1}^2 + 53 h_{0,1}^4 \big),
\end{align}
and, for $j>0$,
\begin{align}
C_j^{(0)} & = 0,\\
C_j^{(1)} & = -\frac{N_{00}^{\rm{sp}}}{N_{j0}^{\rm{sp}}} \left(\frac{1}{j \sqrt{2 \pi}}\right), 
\\ 
C_j^{(2)} & = -\frac{N_{00}^{\rm{sp}}}{N_{j0}^{\rm{sp}}} \left(\frac{1}{j \sqrt{2 \pi}}\right)^2 \left(1 - j h_{0,1}\right),
\end{align}
and
\begin{align}
C_j^{(3)} & = -\frac{N_{00}^{\rm{sp}}}{N_{j0}^{\rm{sp}}} \left(\frac{1}{j \sqrt{2 \pi}}\right)^3 \!\! \bigg[1 - 2 j h_{0,1}
\\ \nonumber
& \qquad +\frac{j^2}{4}\left(3h_{0,2}+7h_{0,1}^2\right)\bigg].
\end{align}
The $h_{n,p}$ are defined in Eq.~\eqref{eq_Hnp}.
Using the notation introduced in Eq.~(\ref{eq_condfracgamma}),
we find
\begin{align}
\label{eq_gamma22}
\gamma_2^{(2)}=
-2 C_0^{(2)}
- \frac{3}{8\pi}\;_4F_3(1,1,1,5/2,2,2,2,1/4) ,
\end{align}
\begin{eqnarray}
\label{eq_gamma213}
\gamma_{2,1}^{(3)}=-
\frac{3}{4(2 \pi)^{3/2}}\;_5F_4(1,1,1,1,5/2,2,2,2,2,1/4)\nonumber \\
+\frac{h_{01}^3}{6(2 \pi)^{3/2}} 
+\frac{h_{01}h_{02}}{2(2 \pi)^{3/2}}
+\frac{h_{03}}{3 (2 \pi)^{3/2}},
\end{eqnarray}
\begin{eqnarray}
\label{eq_gamma223}
\gamma_{2,2}^{(3)}= \sum_{j=1}^{\infty} \sum_{k=1}^{\infty} 
\frac{2^{1/2-2j-2k}\Gamma(j+k+3/2)}{4 j k(j+k)\pi^2 \; j! \; k!} ,
\end{eqnarray}
\begin{eqnarray}
\label{eq_gamma233}
\gamma_{2,3}^{(3)}=
-\frac{1}{\sqrt{2 \pi}} h_{0,1}
\gamma_2^{(2)},
\end{eqnarray}
and
\begin{eqnarray}
\gamma_2^{(4)} = 0.406786416075. 
\end{eqnarray}
In Eqs.~(\ref{eq_gamma22}) and (\ref{eq_gamma213}),
$_q F_p$ denotes the generalized hypergeometric function.
The numerical values of $\gamma_2^{(2)}$, $\gamma_{2,1}^{(3)}$,
$\gamma_{2,2}^{(3)}$ and $\gamma_{2,3}^{(3)}$
are listed in Table~\ref{tab_coefficient}.

The approach discussed above can be extended
to account for the effective range dependence
of the condensate fraction, yielding the result discussed
in the last paragraph of Sec.~\ref{sec_twobody}.


\begin{thebibliography}{10}
\bibitem{notation1}
The generalized harmonic number $H_{n,p}$ can be written as
$\psi^{(p-1)}(n+1)=(-1)^p \Gamma(p) [\zeta(p)-H_{n,p}]$
for $p>1$ and $\psi^{(0)}(n+1)= -\gamma + H_{n,1}$ for $p=1$,
where $\zeta()$ denotes 
the zeta function,
$\Gamma()$ 
the gamma function,
$\psi^{(p)}()$
the polygamma function 
and
$\gamma$ Euler's constant.
\bibitem{notation2}
The solid and dashed lines terminate at 
$a_{s}(0) \approx 0.006 a_{{\rm ho}}$
since the solution to the trapped two-body
problem with a hardcore like Gaussian potential
via the employed B-spline approach becomes
numerically challenging as $a_{s}/a_{{\rm ho}}$ becomes
large.
\end{thebibliography}
\end{document}